\documentclass[aps,prx,twocolumn,amsmath,amssymb,notitlepage,noeprint,nofootinbib,superscriptaddress,longbibliography]{revtex4-1}
\pdfoutput=1
\usepackage[latin9]{inputenc}
\usepackage{braket}
\usepackage{xcolor}
\usepackage{verbatim}
\usepackage{graphicx}
\usepackage{esint}
\usepackage{float}
\usepackage{bm}
\usepackage{notes2bib}
\definecolor{darkblue}{rgb}{0,0,0.6}
\usepackage[colorlinks,linkcolor=darkblue,citecolor=darkblue,urlcolor=darkblue]{hyperref}

\let \a=\alpha
\newcommand{\ta}{\tau_{\alpha}}
\newcommand{\beq}{\begin{equation}} \newcommand{\eeq}{\end{equation}}

\newcommand{\rev}[1]{{\color{black} {#1}}}

\begin{document}

\title{Thirty milliseconds in the life of a supercooled liquid}

\author{Camille Scalliet}

\affiliation{Department of Applied Mathematics and Theoretical Physics, University of Cambridge, Wilberforce Road, Cambridge CB3 0WA, United Kingdom}

\author{Benjamin Guiselin}

\affiliation{ENSL, CNRS, Laboratoire de physique, F-69342 Lyon, France}

\author{Ludovic Berthier}

\affiliation{Laboratoire Charles Coulomb (L2C), Universit\'e de Montpellier, CNRS, 34095 Montpellier, France}

\affiliation{Yusuf Hamied Department of Chemistry, University of Cambridge, Lensfield Road, Cambridge CB2 1EW, United Kingdom}

\date{\today}

\begin{abstract}
We combine the swap Monte Carlo algorithm to long multi-CPU molecular dynamics simulations to analyse the equilibrium relaxation dynamics of model supercooled liquids over a time window covering ten orders of magnitude for temperatures down to the experimental glass transition temperature $T_g$. The analysis of \rev{several} time correlation functions coupled to spatio-temporal resolution of particle motion allow us to elucidate the nature of the equilibrium dynamics in deeply supercooled liquids. We find that structural relaxation starts at early times in rare localised regions characterised by a waiting time distribution that develops a power law near $T_g$. At longer times, relaxation events accumulate with increasing probability in these regions as $T_g$ is approached. This accumulation leads to a power-law growth of the linear extension of relaxed domains with time with a large, \rev{temperature-dependent} dynamic exponent. Past the average relaxation time, unrelaxed domains slowly shrink with time due to relaxation events happening at their boundaries. Our results provide a complete microscopic description of the particle motion responsible for key experimental signatures of glassy dynamics, from the shape and temperature evolution of relaxation spectra to the core features of dynamic heterogeneity. They also provide a microscopic basis to understand the emergence of dynamic facilitation in deeply supercooled liquids and allow us to critically reassess theoretical descriptions of the glass transition. 
\end{abstract}

\maketitle

\section{Introduction}

\label{sec:introduction}

There exists a large corpus of experimental studies analysing the physical properties of supercooled liquids undergoing a glass transition~\cite{ediger1996supercooled,doi:10.1063/1.1286035,angell1995formation}. Glassy systems exhibit well-established signatures characterising their thermodynamic, rheological, and dynamic properties. An important goal of this experimental quest is to develop a \rev{sufficiently} precise understanding of the physical behaviour of liquids undergoing a glass transition to guide and constrain theoretical developments. A successful theoretical framework should explain the observed behaviours with precise assumptions that can be directly tested by experiments~\cite{Berthier2011,debenedetti2001supercooled}. This program is not yet complete, and different theoretical explanations remain able to account for experimental results using \rev{hypotheses} that \rev{can be} difficult to validate experimentally~\cite{tarjus2011overview,lubchenko2007theory,chandler2010dynamics,tarjus2005frustration,royall2015role}.   

Computer simulations have an important role to play in this endeavour as they offer by construction a complete spatio-temporal resolution of glassy dynamics, and the possibility to measure observables which are difficult or impossible to access experimentally~\cite{barrat2022computer,berthier2022modern}. For a long time, a major obstacle was the inability to study realistic models of glassy liquids in the temperature regime relevant to experiments. The situation changed radically five years ago when the swap Monte Carlo algorithm~\cite{Grigera2001} was optimised and novel glass models were developed~\cite{ninarello2017models,berthier2019efficient,parmar2020stable}. The swap Monte Carlo algorithm employs unphysical particle moves to accelerate the equilibration of supercooled liquids and can reach equilibrium states down to the experimental glass transition temperature $T_g$ or even below. This algorithmic development allowed progress regarding the analysis of structural and thermodynamic properties of liquid states~\cite{berthier2017configurational,berthier2019zero,PhysRevE.102.042129,guiselin2022statistical,nishikawa2022relaxation}, as well as characterisation of the glass below $T_g$~\cite{scalliet2017absence,Ozawa2018,Scalliet2019,khomenko2020depletion,wang2019low,PhysRevLett.124.225502,wang2019sound}.

However, because it employs unphysical particle motion, the swap Monte Carlo algorithm cannot be used to directly analyse the dynamics of supercooled liquids near $T_g$. To date, simulations of the dynamics were mostly performed in a relatively high temperature regime corresponding, at best, to an average relaxation time up to \rev{one microsecond}, when converted into experimental units~\cite{eastwood2013rotational,simons,howtomeasure}. This corresponds to a temperature scale near the mode-coupling temperature crossover $T_{\rm mct} > T_g$. Fortuitously, experiments performed with colloidal particles cover a similar dynamic range~\cite{Hunter_2012,doi:10.1080/00018732.2016.1200832}. As a result, a detailed microscopic characterisation of the dynamics in the mode-coupling regime $T \geq T_{\rm mct}$ is available~\cite{kob1995testing,kob1999computer}. There exists ample evidence that physics in the regime $T<T_{\rm mct}$ may be of a different nature~\cite{PhysRevLett.102.085703,kob2012non,Dalle-Ferrier2007,das2022crossover,coslovich2018dynamic} but a thorough numerical exploration of glassy dynamics deep in this temperature regime is currently lacking.   

Here we show that month-long multi-CPU molecular dynamics simulations started from configurations that are first equilibrated using the swap Monte Carlo algorithm open a novel window to analyse the equilibrium dynamics of deeply supercooled liquids in the regime $T_g <  T < T_{\rm mct}$. In practice, we follow the \rev{equilibrium dynamics} over ten decades in time with particle-scale resolution at temperatures down to $T_g$. Converted to experimental units, this approach allows us to follow the entire structural relaxation at temperatures well below the mode-coupling crossover $T < T_{\rm mct}$, up to relaxation times of $\tau_\alpha \approx 10$~ms. For even lower temperatures $T \approx T_g$, we study the first ten decades of the structural relaxation up to a maximal timescale of about 30 milliseconds for our longest simulations. 

We explore the relaxation dynamics of model supercooled liquids in a regime that was too difficult or impossible to access before. This \rev{important} numerical effort has led to a previous work~\cite{guiselin2022microscopic}, where we concentrated on the emergence of excess wings in dynamic spectra obtained at low temperatures. Our ambition and focus here are very different as we provide a complete view of all relaxation processes from microscopic times up to timescales several times longer than the relaxation time.

Our philosophy in this article is to first report our numerical observations and quantify them with as little interference from theoretical models as possible. We pay special attention to the regime below $T_{\rm mct}$ that has not been accessed before and contrast our findings with earlier work at higher temperatures. In the final part of the manuscript only, we critically compare our results to existing theoretical frameworks. For the two models studied, a clear physical picture of the structural relaxation emerges at low temperatures, which stems from early relaxation events that are broadly distributed \rev{followed by} increasingly correlated motions in space and time. These observations account for the emergence of dynamic facilitation in the dynamics of deeply supercooled states. 

Our manuscript is organised as follows. 
In Sec.~\ref{sec:methods} we define our computer models and numerical strategy.
In Sec.~\ref{sec:ensemble} we present results concerning ensemble-averaged time correlation functions.
In Sec.~\ref{sec:visualisation} we offer visualisation of the relaxation dynamics over a broad range of timescales, lengthscales, and temperatures.
In Sec.~\ref{sec:clusters} we analyse in more detail the early times of the relaxation.
In Sec.~\ref{sec:towards} we explain how the structural relaxation unfolds from early to large times.
In Sec.~\ref{sec:discussion} we provide a discussion of our results, comparing them with earlier numerical work and theoretical views.  

\section{Computer models and methods}

\label{sec:methods}

\subsection{Glass-forming models}

\label{sub:models}

We study size-polydisperse mixtures of soft repulsive spheres in two and three spatial dimensions, $d=2,3$. These two models have been shown to be representative computational glass-formers~\cite{berthier2019zero,berthier2017configurational} and extensively studied before~\cite{doi:10.1063/1.5113477,PhysRevLett.124.225502,Ozawa2018,wang2019low}. The particle diameters $\{ \sigma_i \}$ are drawn from the probability distribution $\mathcal{P}(\sigma) =\mathcal{A}/\sigma^3$ with $\mathcal{A}$ a normalisation constant, with bounds $\sigma_\mathrm{min} \leq \sigma_i \leq \sigma_\mathrm{max}$. Two particles $i$ and $j$ at positions $\bm{r}_i$ and $\bm{r}_j$ and separated by a distance $r_{ij} = |\bm{r}_{ij}|=|\bm{r}_i - \bm{r}_j|$ interact within a cutoff $r_{ij}/\sigma_{ij} < x_c=1.25$ with a repulsive interaction \rev{potential}:
\begin{equation}
v(r_{ij})=\epsilon\left(\frac{\sigma_{ij}}{r_{ij}}\right)^{12}+c_0 + c_2\left(\frac{r_{ij}}{\sigma_{ij}}\right)^2+ c_4\left(\frac{r_{ij}}{\sigma_{ij}}\right)^4,
\end{equation}
where the constants $c_0=-28\epsilon/x_c^{12}$, $c_2=48\epsilon/x_c^{14}$, and $c_4=-21\epsilon/x_c^{16}$ ensure continuity of the potential and of its first two derivatives at the cutoff $x_c$. We employ a non-additive mixing rule $\sigma_{ij} =0.5(\sigma_i+\sigma_j)( 1-\eta|\sigma_i-\sigma_j|)$ to avoid fractionation and crystallization at low temperatures~\cite{ninarello2017models}. All particles have an equal mass $m$. We use the average diameter $\overline \sigma$ as unit length, $\epsilon$ as unit energy \rev{with the Boltzmann constant set to unity} and $\sqrt{m\overline{\sigma}^2/\epsilon}$ as unit time. In these units, we choose $\sigma_\mathrm{min} = 0.73$, $\sigma_\mathrm{max} = 1.62$ and $\eta = 0.2$.

We simulate $N$ particles at number density $\rho=N/L^d = 1$ in a cubic or square box of linear size $L$ with periodic boundary conditions. The results reported for the $3d$ model are obtained for systems of $N=1200$ particles. We have performed simulations of a larger system with $N=10^4$ particles to check for finite-size effects in the dynamics and for visualisation purposes. Measurements in the $2d$ model are reported for $N=2000$ particles. Some $2d$ simulations were performed with $N=10^4$ particles as well, in particular to generate snapshots. 

\subsection{Monte Carlo and molecular dynamics simulations}

\label{sub:nummethods}

We use a hybrid scheme combining Monte Carlo (MC) particle-swap moves and molecular dynamics (MD) in order to generate equilibrium configurations~\cite{berthier2019efficient}. The algorithm alternates between ordinary MD simulation sequences during which the particle positions and velocities evolve with a fixed particle diameter, and Monte Carlo sequences during which particles diameters are swapped at fixed positions and velocities. The MD sequences last $t_{\rm MD}$ and take place at constant temperature $T$ imposed by a Nos\'e-Hoover thermostat. To perform MC sequences, the positions and velocities are frozen and $n_{\textrm{swap}}N$ swap moves are attempted. During a swap move, two particles are randomly selected and an exchange of their diameters is proposed. After computing the change in potential energy, the move is accepted following the Metropolis acceptance rule at temperature $T$, which ensures equilibrium sampling. Following Ref.~\cite{berthier2019efficient} we optimise $t_{\rm MD}$ and $n_{\rm swap}$ for maximal efficiency. Using this hybrid algorithm, we prepare a large number of statistically independent configurations $\mathcal{N} \in [100,\ 500]$ for temperatures ranging from the onset of glassy dynamics $T_o$ down to the experimental glass transition temperature $T_g$. These characteristic temperature scales are defined and numerically determined below.

We take these equilibrium configurations as initial conditions for standard MD simulations without swap MC moves. We perform microcanonical $NVE$ simulations in $3d$ and canonical $NVT$ simulations with a Nos\'e-Hoover thermostat in $2d$, using in both cases a time step $dt=0.01$ for the numerical integration of the equations of motion. \rev{These different choices are made for convenience only, as the statistical ensemble does not influence the local dynamics.} The duration of most simulations is $10^{7}$, corresponding to $10^9$ MD steps and a simulation time of about one week. Ensemble-averaged observables are computed by averaging over $\mathcal{N}$ independent trajectories. For selected state points in $2d$ we increase the simulation time to $8.2 \times 10^{8}$ for $N=10^4$ particles, which was reached by using a parallel code running over 24 processors (CPUs) for two and a half months.

\subsection{Why thirty milliseconds?}

\label{sub:numstrategy}

The temperatures and timescales over which equilibrium dynamics can be probed with MD simulations are limited, \rev{as recently reviewed in Ref.~\cite{simons}.} Molecular dynamics simulations last at most $10^{10}$ MD steps for a simple pair potential simulated over several weeks and a reasonable system size ($N\sim 10^3-10^4$). This translates into a maximum simulation time $t_{\textrm{max}} \sim 10^8$, considering a typical discretisation time step of order $10^{-2}$. \rev{Without the swap Monte Carlo algorithm}, the simulated time is \rev{necessarily} split between equilibration and production runs. To achieve both a proper equilibration and a significant exploration of the configuration space in equilibrium conditions, the MD equilibration run must last at least about \rev{$100$} times the averaged relaxation time $\tau_\alpha$. This sets an upper limit $\ta \leq 10^6$ on the accessible relaxation times even using extensive simulations and an efficient implementation of the MD. Most numerical studies have therefore investigated the dynamics of supercooled liquids with $\tau_\alpha \leq 10^5$.

Our numerical strategy completely circumvents the need for time consuming equilibration runs by exploiting the huge equilibration speedup afforded by the swap Monte Carlo algorithm. This idea is in its infancy~\cite{howtomeasure,PhysRevLett.127.088002,guiselin2022microscopic}. In this approach, preparing multiple equilibrated configurations down to $T_g$ is a simple task, and we can thus ensure both equilibration and statistical accuracy of the results. This implies that the computational time can be entirely spent on the production runs to simulate the dynamics of low temperature states over a time window limited to $t_{\textrm{max}} \approx 10^8$. Crucially~\cite{howtomeasure}, this time window is available even at temperatures where $\ta \gg t_{\textrm{max}}$, which were inaccessible in equilibrium conditions in previous work.

Following earlier work~\cite{ninarello2017models,doi:10.1063/1.5113477,howtomeasure}, we translate simulation timescales into experimental ones using the structural relaxation time $\tau_o$ at the onset temperature $T_o$ as reference timescale. For a broad range of molecular liquids~\cite{PhysRevE.86.041507}, one measures $\tau_o \approx 10^{-10}$~s. For the computer models studied here, one finds $\tau_o\approx 3$ in simulation units. The production runs over which ensemble-averaged quantities are measured last $10^7 \approx 3 \times 10^6\,\tau_o$ in simulation units, which translates into $3 \times 10^6\,\tau_o \approx 0.3$~ms in experimental units. Below, we shall also present results of very long simulations which last $t_{\textrm{max}} = 8.2 \times 10^{8}$. In experimental units, this corresponds to observing the relaxation dynamics of a supercooled liquid over about $30$~ms, \rev{as announced.}

\section{Equilibrium relaxation dynamics down to the experimental glass transition}

\label{sec:ensemble}

\subsection{Time correlation functions}
\label{sub:time}

We first characterise the equilibrium dynamics using ensemble-averaged time correlation functions. In $3d$, we use the self-intermediate scattering function
\begin{equation}
F_s(t)=\left\langle \frac{1}{N}\sum_{i=1}^N\cos\left[\bm{q}\cdot \delta \bm{r}_i(t)\right]\right\rangle, 
\label{eq:fs}
\end{equation}
where $\delta \bm{r}_i(t) = \bm{r}_i(t)-\bm{r}_i(0)$, and the brackets indicate an average over the $\mathcal{N}$ independent runs at temperature $T$. We also perform an angular average over wavevectors with $|\bm{q}|=6.9$, corresponding to the first peak in the total structure factor $S(q)$. We define the $\alpha$-relaxation time $\ta(T)$ as $F_s(\ta) = e^{-1}$.

\begin{figure}
    \includegraphics[width=\columnwidth]{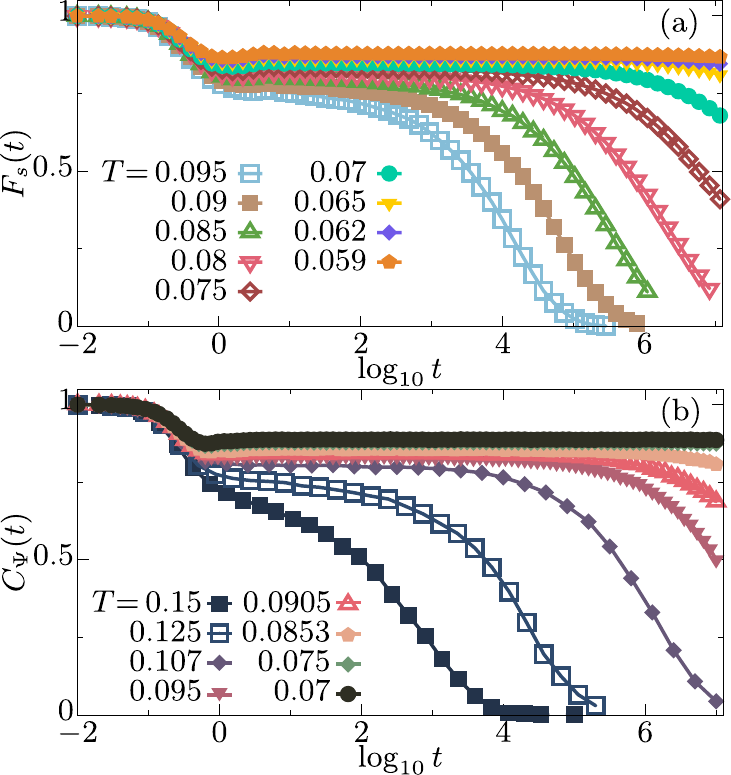}
    \caption{{\bf Equilibrium time correlation functions characterising relaxation dynamics of deeply supercooled liquids.} (a) Self-intermediate scattering function $F_s$ in $3d$. (b) Bond-orientational correlation function $C_\Psi$ in $2d$. In both panels, we only show temperatures between the mode-coupling crossover and the experimental glass transition temperature, $T_{g} \leq T \leq T_{\rm mct}$.}
    \label{fig:timecorrel}
\end{figure}

In Fig.~\ref{fig:timecorrel}(a), we present equilibrium results for $F_s(t)$ in $3d$ at several temperatures. We concentrate on data in the unexplored low-temperature regime where $\ta > 10^4$. The highest temperature shown, $T=0.095$, roughly corresponds to the lowest temperature explored in the majority of earlier studies while the lower temperature shown is close to the experimental glass transition temperature $T_g$. This broad temperature range concretely demonstrates how our numerical strategy opens a novel temperature window to study the dynamical slowdown in supercooled liquids.

All correlation functions display a fast decay at very short time $t \approx 1$ due to thermal motion within the arrested amorphous structure. At very low temperature, elasticity and inertial dynamics give rise to a weak oscillatory decay towards the plateau. After an extended plateau regime which becomes more extended at lower temperature, full decorrelation is eventually observed taking the form of a stretched exponential. As temperature decreases, structural relaxation shifts to longer times and it is no longer observed for temperatures $T \lesssim 0.075$. At the lowest temperature $T \lesssim 0.059$, we observe an extended plateau covering about 7 decades in time. More careful inspection reveals that the plateau is not \rev{strictly} constant but decays extremely weakly with time, as we shall illustrate more clearly below. Recall that despite the absence of decorrelation in the numerical time window, all correlation functions are representative of equilibrium dynamics. 

In $2d$, collective long-ranged fluctuations give rise to particle displacements which affect the behaviour of $F_s(t)$ \rev{in an unwanted way~\cite{illing2017mermin,vivek2017long}}. We measure the bond-orientational correlation function $C_{\Psi}(t)$ which is sensitive to changes in the local environment of particles~\cite{flenner2015fundamental,flenner2019viscoelastic}. We define the six-fold order parameter of particle $i$
\begin{equation}
\Psi _i (t) = \frac{1}{n_i(t)} \sum_{j = 1}^{n_i(t)} e^{\mathrm{i} 6\theta _{ij}(t)},
\end{equation}
where $n_i(t)$ is the number of neighbours it has at time $t$, defined as particles $j$ with $r_{ij}<1.45$. The cutoff corresponds to the first minimum in the total pair distribution function, and we have checked that alternative definitions of neighbours via a Voronoi tessellation or a solid-angle based method~\cite{van2012parameter} lead to similar results. The angle $\theta _{ij}(t)$ is defined between the $x$-axis (without loss of generality) and ${\bm r}_{ij}(t)$. The bond-orientational correlation function then reads
\begin{equation}
C_{\Psi} (t) = \left\langle \frac{\sum_i \Psi _i (t) \Psi _i^* (0)}{\sum_i | \Psi _i (0)|^2} \right\rangle ,
\end{equation}
where the star denotes the conjugate complex and the sums run over all particles $i=1\dots N$. We extract the $\alpha$-relaxation time $\ta(T)$ as $C_{\Psi}(\tau_{\alpha}) = e^{-1}$.

In Fig.~\ref{fig:timecorrel}(b), we present equilibrium results for $C_{\Psi} (t)$ in $2d$ at several temperatures, selected as in $3d$ below the mode-coupling crossover. Clearly, the time and temperature evolution of the bond-orientational correlation function is qualitatively similar to that of the self-intermediate scattering function in $3d$ shown in Fig.~\ref{fig:timecorrel}(a). 

\subsection{Relaxation times and temperature scales}

We determine three temperature scales relevant to describe the dynamic slowdown in supercooled liquids: the onset temperature $T_o$, the mode-coupling crossover temperature $T_{\rm mct}$, and the experimental glass transition temperature $T_g$. We first provide the value of these temperature scales, and detail below \rev{how we estimate them.} We obtain $T_o=0.2$, $T_{\rm mct}=0.095$, and $T_g=0.056$ in $3d$, and $T_{o}=0.2$, $T_{\rm mct}=0.12$, and $T_{g}=0.07$ in $2d$. These characteristic temperatures will be useful to interpret and organise our results.

\begin{figure}
    \includegraphics[width=\columnwidth]{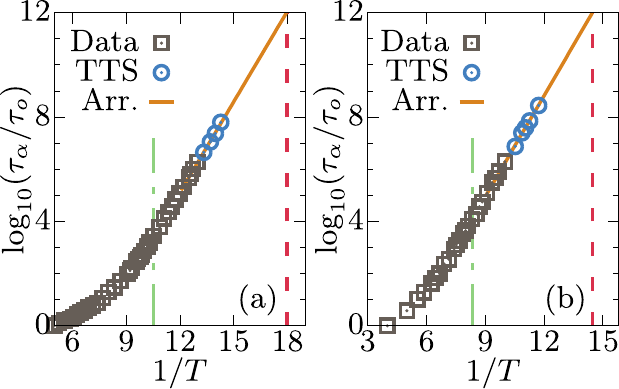}
    \caption{{\bf Averaged relaxation times in $3d$ and $2d$ glass-forming models.} Relaxation time $\tau_\a$, rescaled by its value $\tau_o$ at the onset temperature, as a function of inverse temperature in $3d$ (a) and $2d$ (b). The data points are directly measured from Fig.~\ref{fig:timecorrel}, and extended over $1.5$ decades using time-temperature superposition (TTS). An Arrhenius fit (full line) extrapolates $\ta/\tau_o$ to the value $10^{12}$ to locate the experimental glass transition temperature $T_g$ (dashed line). We concentrate on the regime between $T_{\rm mct}$ (dash-dotted line) and $T_g$.}
    \label{fig:relaxtime}
\end{figure}

In Figs.~\ref{fig:relaxtime}(a,b) we present the relaxation time $\ta$ for $3d$ and $2d$ models, respectively. At the onset temperature of glassy dynamics $T_o$, where the relaxation time equals $\tau_o$, the relaxation time departs from its high-temperature Arrhenius dependence, $\tau_\alpha(T) \propto e^{E_\infty/T}$. The energy scale $E_\infty$ is equal to $E_\infty=0.23$ (in $3d$) and $E_\infty =0.7$ (in $2d$), and $\tau_o \approx 3$ (in $d=2,3$). In Fig.~\ref{fig:relaxtime}, we focus on the supercooled regime and report $\ta$ normalized by $\tau_o$ as a function of the inverse temperature. In this representation, Arrhenius behaviour translates into a straight line. In the simulations, we directly measure the relaxation in the range $\ta < 10^7$, equivalently $\log_{10}(\ta/\tau_o) < 6.5$. These measurements are labeled as `Data' in Fig.~\ref{fig:relaxtime}. 

Then, we locate the mode-coupling crossover temperature $T_{\rm mct}$ by using a power-law fit $\tau_{\alpha}(T) \sim (T - T_{\rm mct})^{-\gamma}$ in the regime $0 \leq \log_{10}(\tau_{\alpha}/\tau_o) \leq 3$~\cite{gotze2009complex}, using the exponents $\gamma=2.7$ (in $2d$) and $\gamma=2.5$ (in $3d$). We find $\ta(T_{\rm mct}) \sim 10^4~\tau_o$ in $2d$ and $3d$.

Direct measurements of $\ta$ stop at temperatures where the correlation functions do not reach $e^{-1}$ in the numerical time window. This occurs roughly 5 decades before the extrapolated glass transition $T_g$, which is defined by $\log_{10}(\ta(T_g)/\tau_o) = 12$. Still, at temperatures where $\ta < 10^7$, the final decay is well-fitted by a stretched exponential $\sim F_0 e^{-(t/\ta)^\beta}$. The stretching exponents $\beta \approx 0.56$ (in $3d$) and  $\beta\approx 0.6$ (in $2d$) are nearly temperature-independent \rev{within statistical uncertainty}, and the amplitude $F_0$ increases weakly with decreasing temperature. This indicates that time-temperature superposition (TTS) is well-obeyed in our systems. Fixing $\beta$ to a constant value, we can use TTS to estimate $\ta$ when it falls outside our numerical window~\cite{howtomeasure}. In practice, we can safely extend our measurements of $\ta(T)$ by an additional $1.5$ decades, as indicated by the `TTS' points in Fig.~\ref{fig:relaxtime}, which then reach $\log_{10} (\ta / \tau_o) \approx 8$ ({\it i.e.}, about 10~ms).

To estimate the experimental glass transition temperature $T_g$, we must extrapolate the relaxation time data by 4 additional orders of magnitude. To do so, we describe $\ta$ over the remaining decades using an Arrhenius fit $\tau_\alpha(T) \propto e^{E_A/T}$, with $E_A = 2.67$ (in $3d$) and $E_A = 2.97$ (in $2d$), and locate $T_g$ where $\log_{10}(\ta(T_g)/\tau_o)=12$. Although the temperature dependence of $\ta$ is not purely Arrhenius over the entire numerical window, our choice of an Arrhenius extrapolation which neglects fragility provides a lower bound to the correct value of $T_g$. 

The swap Monte Carlo algorithm allows us to easily prepare equilibrium configurations even at temperatures close to the determined $T_g$. We can thus safely claim that we analyse the equilibrium dynamics of liquids down to the experimental glass temperature $T_g$. The curves at the lowest temperatures shown in Fig.~\ref{fig:timecorrel} correspond to the dynamics at, or very close to, the experimental glass transition temperature where we can access the first 30~ms of the relaxation dynamics in our longest simulations.

\subsection{Mean-squared displacements: E pur si muove!}

\label{sub:msd}

\begin{figure}
  \includegraphics[width=\columnwidth]{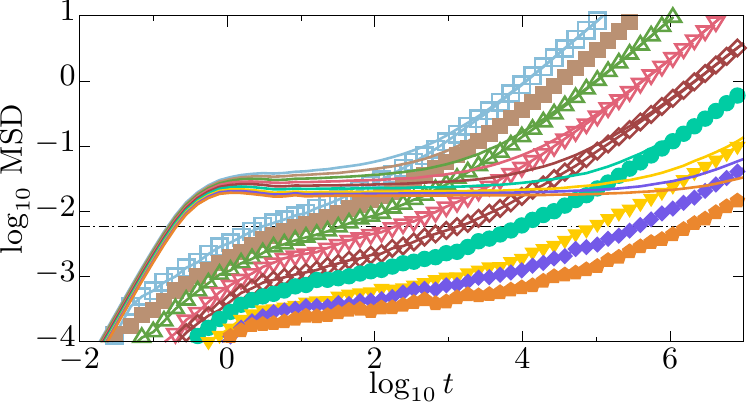}
\caption{{\bf Mean-squared displacement in $3d$ deeply supercooled states}. From $T_{\rm mct}$ to $T_g$ (left to right) in the normal dynamics $\Delta(t)$ (full lines) and in the inherent structures (IS) $\Delta^{\rm IS}(t)$ (symbols). The legend is as in Fig.~\ref{fig:timecorrel}(a). The dash-dotted line corresponds to $\Delta=6 \times 10^{-3}$, used in Fig.~\ref{fig:vanhove} below.}
\label{fig:disp_3d}
\end{figure}

To characterise the average motion of the particles we compute the mean-squared displacement (MSD):
\begin{equation}
\Delta(t) = \left\langle \frac{1}{N}\sum_{i=1}^N \big|\bm{r}_i(t)-\bm{r}_i(0)\big|^2\right\rangle.
\label{eq:MSD}
\end{equation}
The full lines in Fig.~\ref{fig:disp_3d} present the time dependence of the MSD at various temperatures in $3d$. \rev{Data are not presented in $2d$ because of the large and spurious collective fluctuations mentioned above.} When temperature is not too low, the curves exhibit the usual time dependence with a ballistic regime at early times, followed by a long plateau at intermediate times before finally entering a diffusive regime at very long times which we can observe down to $T=0.09$. As the temperature decreases, the intermediate plateau extends over longer timescales, and the diffusive regime eventually shifts outside the numerical time window. Close to $T_g$, we can only observe the very beginning of the escape from the plateau, mirroring the behaviour of the self-intermediate scattering function in Fig.~\ref{fig:timecorrel}(a). The MSD at long times reaches the small value $\Delta \approx 0.03$ corresponding to an average particle displacement of about one fifth of the average particle diameter. In other words, the system appears totally frozen on the very long timescale explored by the simulations.  And yet, as we shall see, particles move!

Indeed, a more subtle picture emerges by filtering thermal motion~\cite{schroder2000crossover,dyremsdis}. We use a conjugate gradient algorithm to bring the configurations explored dynamically $\lbrace \bm{r}_i(t)\rbrace_{i=1..N}$ to their nearest potential energy minimum, also called inherent state (IS), $\lbrace \bm{r}_i^{\rm IS}(t)\rbrace_{i=1..N}$. From these, we compute a version of the MSD, $\Delta^{\rm IS}(t)$, where the erratic thermal motion of particles exploring their cages no longer contributes. When particles simply fluctuate around their average position, this procedure returns a vanishing value for $\Delta^{\rm IS}$, while $\Delta$ plateaus at the Debye-Waller factor.

The data for $\Delta^{\rm IS}$ shown in Fig.~\ref{fig:disp_3d} paint a completely different picture. The two MSD curves only coincide in the long-time diffusive regime for large particle displacements, where the small contribution due to thermal vibrations becomes negligible. At early times in the ballistic regime, we have $\Delta^{\rm IS} \ll \Delta$ as previously reported~\cite{dyremsdis}. Surprisingly, we observe that close to $T_g$, $\Delta^{\rm IS}(t)$ develops a non-trivial time dependence as soon as $t > 1$. Although the average structural relaxation time is close to $\ta = 10^{12}$, there are already non-trivial particle motion and relaxation taking place at times that are many orders of magnitude shorter. We conclude that the extended plateau in the MSD in fact masks early non-trivial relaxation events which take place much before the average structural relaxation time. \rev{These motions would be difficult to detect when using the self-intermediate scattering function, as pioneered in Ref.~\cite{schroder2000crossover}. The non-trivial time evolution of the MSD when measured in inherent structures also shows that a large number of IS are explored over the time window preceding the structural relaxation, which raises questions about the relevance of the potential energy landscape to describe the $\a$-relaxation dynamics of deeply supercooled liquids.}

\rev{These observations constitute} a central theme of this work: by resolving particle motion in equilibrium conditions, we reveal how the structural relaxation unfolds over an extended period of time between $t \approx 1$ and $t \gg \ta$ \rev{at very low temperatures.} In the rest of the article, we analyse the corresponding particle motion and develop a physical understanding of the structural relaxation in deeply supercooled liquids.  

\section{Qualitative overview of the relaxation dynamics}

\label{sec:visualisation}

\subsection{Direct visualisation of the relaxation}

To resolve relaxation dynamics at the particle scale, we use the bond-breaking correlation function $C_B^i(t)$~\cite{yamamoto1998heterogeneous,shiba2012relationship}. By definition, this quantity measures the fraction of the neighbours of particle $i$ defined at time $t=0$ that are still neighbour later at time $t$
\begin{equation}
C_B^i(t)=\frac{n_i(t|0)}{n_i(0)},
\label{eq:cb}
\end{equation}
where $n_i(0)$ is the number of neighbours of particle $i$ at time $t=0$, and $n_i(t|0)$ the number of those particles that are still neighbour after time $t$. At $t=0$, neighbours are defined via the criterion $r_{ij}/\sigma_{ij}<1.485$ (in $3d$) or $r_{ij}/\sigma_{ij}<1.3$ (in $2d$), which corresponds to the first minimum in the rescaled pair distribution function constructed using distances $r_{ij}$ rescaled by $\sigma_{ij}$. At $t>0$, we define neighbours via $r_{ij}/\sigma_{ij}<1.7$ (in $d=2,3$) which is below the second peak in the rescaled pair distribution function. This slightly larger cutoff avoids spurious noise due to particles frequently exiting or entering the neighbour shell due to thermal vibrations~\cite{flenner2019viscoelastic}. 

With these choices, $C_B^i(t)$ \rev{only decays whenever} a rearrangement involving particle $i$ takes place, and is thus an excellent indicator of local relaxation events~\cite{yamamoto1998heterogeneous}\rev{, although not being a single-particle quantity. Thresholding single-particle displacements (physical displacements or the ones computed from the visited inherent states) yields qualitatively similar pictures, but the definition of local relaxation may become blurred, especially at the highest studied temperatures.} The correlation $C_B^i(t)$ thus tracks the cumulative effect of rearrangements taking place around particle $i$ over time. Particles which have lost half of their initial neighbours, $C_B^i \leq 0.5$, are called `mobile'~\cite{widmer2009localized}. Particles with $C_B^i=1$ have not rearranged and are called `quiescent'. We define 
\begin{equation}
    C_B(t) = \left\langle \frac{1}{N} \sum_{i=1}^N C_B^i (t)\right\rangle
\end{equation}
as the ensemble-average of $C_B^i(t)$.

\begin{figure}
  \includegraphics[width=\columnwidth]{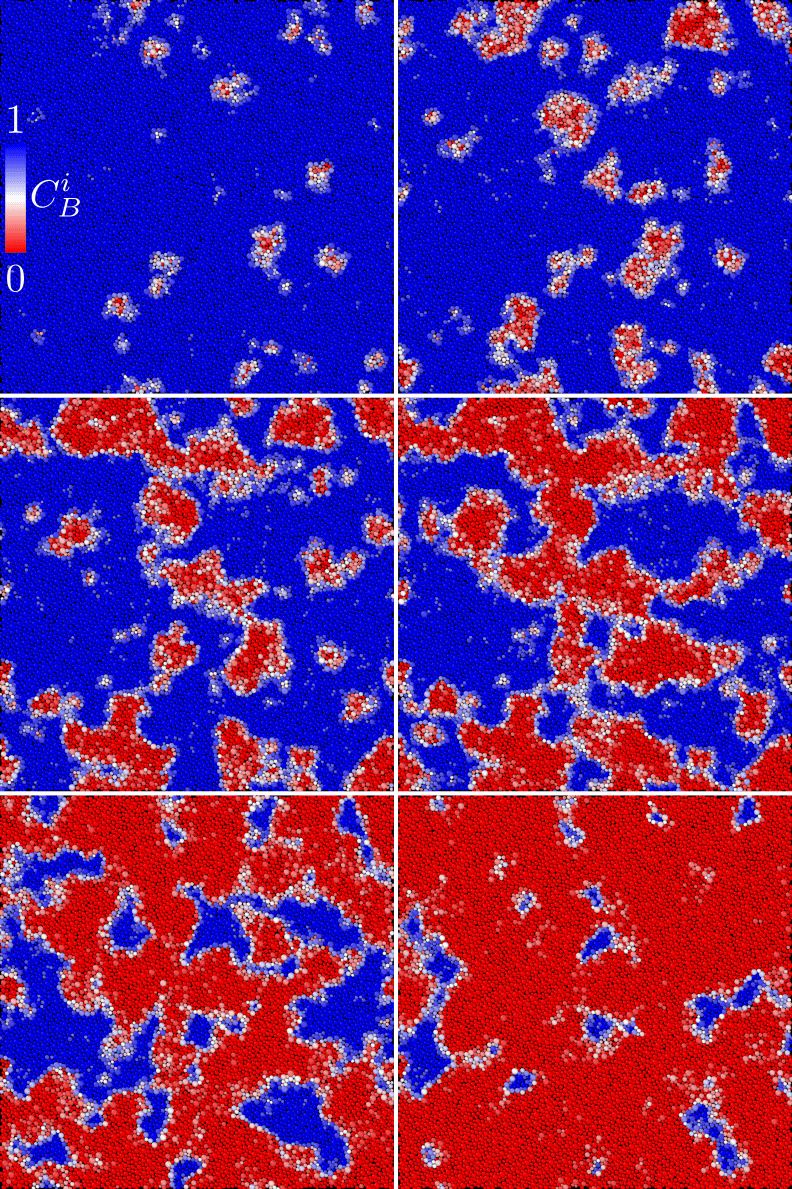}
\caption{
{\bf Thirty milliseconds in the life of a supercooled liquid}. Spatio-temporal evolution (left to right, top to bottom) of the structural relaxation in a $2d$ deeply supercooled liquid at $T=0.09$ where $\ta = 1.5 \times 10^8$. 
Particle colours indicate the bond-breaking correlation $C_B^i(t_j )$, from blue (quiescent) to red (mobile), \rev{following the colour bar indicated top left}. The snapshots are taken at times $t_1,\dots, t_6 = 2 \times 10^6$, $10^7$, $4.6\times 10^7$, $1.1\times 10^8$, $2.3\times 10^8$, $8.2\times 10^8$, at which the average bond-breaking correlation equals $0.97$, $0.86$, $0.7$, $0.54$, $0.33$, and $0.1$. The linear size of the box is $L=100$.} 
\label{fig:redblue}
\end{figure} 

We illustrate the spatio-temporal evolution of the relaxation dynamics in a two-dimensional liquid in Fig.~\ref{fig:redblue}. \rev{The corresponding movies are provided in Supplementary Material~\cite{SMmovies}.} The temperature is $T = 0.09$, well below $T_{\rm mct}=0.12$. This temperature is the lowest at one at which we can observe full decorrelation within the simulation time $t_{\rm max}= 8.2 \times 10^8$, equivalent to $30$~ms, since the relaxation time defined by $C_\Psi(\ta)=e^{-1}$ is $\ta = 1.5 \times 10^8$. On a logarithmic scale, this temperature is thus roughly halfway between $T_{\rm mct}$ ($\ta/\tau_o \approx 10^4$) and $T_g$ ($\ta/\tau_o \approx 10^{12}$). We show snapshots of the liquid at six times along the trajectory $t_1,\dots, t_6 = 2\times 10^6$, $10^7$, $4.6\times 10^7$, $1.1\times 10^8$, $2.3\times 10^8$, $8.3\times 10^8$. The particles are coloured according to the value of the bond-breaking correlation at time $t_j$ using a blue to red code for $C_B^i(t_j)=1$ to $0$. 

At time $t_1$, two decades before $\ta$, the average correlation $C_B(t_1) =0.97$ is still very close to one. While most particles are quiescent (blue), we distinguish a few mobile particles (white/red), clustered into sparse, compact and localised domains. These clusters typically contain a few red particles, which have $C_B^i(t_1)=0$. These particles have undergone a series of nontrivial rearrangements which led to a complete renewal of their local environment. Particles at the boundary of these clusters typically appear white, indicating $C_B^i(t_1)=0.5$: they have lost half of their initial neighbours (those inside the mobile cluster). The images display the raw data for $C_B^i$, with no spatial averaging. The sharp separation observed between mobile and quiescent particles is thus a genuine physical effect. When the average correlation is $C_B=0.5$ near time $t_4$, we see that in real space most particles have either $C_B^i=1$ or $C_B^i=0$ with equal probability. This important property becomes very clear at this very low temperature, as we demonstrate shortly. 

At time increases in Fig.~\ref{fig:redblue}, we observe that relaxation proceeds via two distinct mechanisms. From one snapshot to the next, we first observe the appearance of new relaxed clusters in regions that were quiescent in the preceding snapshot. Clearly, the emergence of such new clusters is a stochastic process characterised by a time distribution that we analyse in Sec.~\ref{sec:waiting}.

The second physical process leading to the appearance of newly relaxed particles is via the coarsening of mobile clusters which exist at time $t_j$ and get larger at time $t_{j+1}$. The time dependence of the domain size is studied in Sec.~\ref{sec:coarsening}. At large times, the growth of relaxed regions leads to their coalescence, so that the system exhibits a bicontinuous structure of mobile/quiescent domains near $\ta$. This bicontinuous structure is obviously characterised by a distribution of domain sizes that we investigate in Sec.~\ref{sec:coarsening}. 

At very long times in Fig.~\ref{fig:redblue} we observe small regions that are still quiescent after $5.5\,\ta$. They represent the most stable regions of the structure at time $t=0$. Compared to the substantial past and present research focusing on fast relaxing regions and defects, very little is known about these slow domains. Inspection of the last two snapshots reveals that the slow domains are slowly invaded by the red mobile regions and slowly shrink in size as time increases. In Sec.~\ref{sec:lifetime} we discuss the physics associated to these slow domains, and their relation with the lifetime of dynamic heterogeneities.

\subsection{Evolution with temperature}

The physical picture of the structural relaxation revealed by the images in Fig.~\ref{fig:redblue} appears relatively simple because there is a very sharp separation between mobile and quiescent particles. However, such a clear distinction only emerges when temperature is low enough, and is not present in the temperature scale traditionally studied in computer simulations. To illustrate this point, we compare the pattern of structural relaxation at different temperatures. \rev{The corresponding movies are provided in Supplementary Material~\cite{SMmovies}.} We keep the value of the average correlation function $C_B(t)$ equal, as this corresponds to a similar degree of decorrelation from the initial structure. 

\begin{figure}
    \includegraphics[width=\columnwidth]{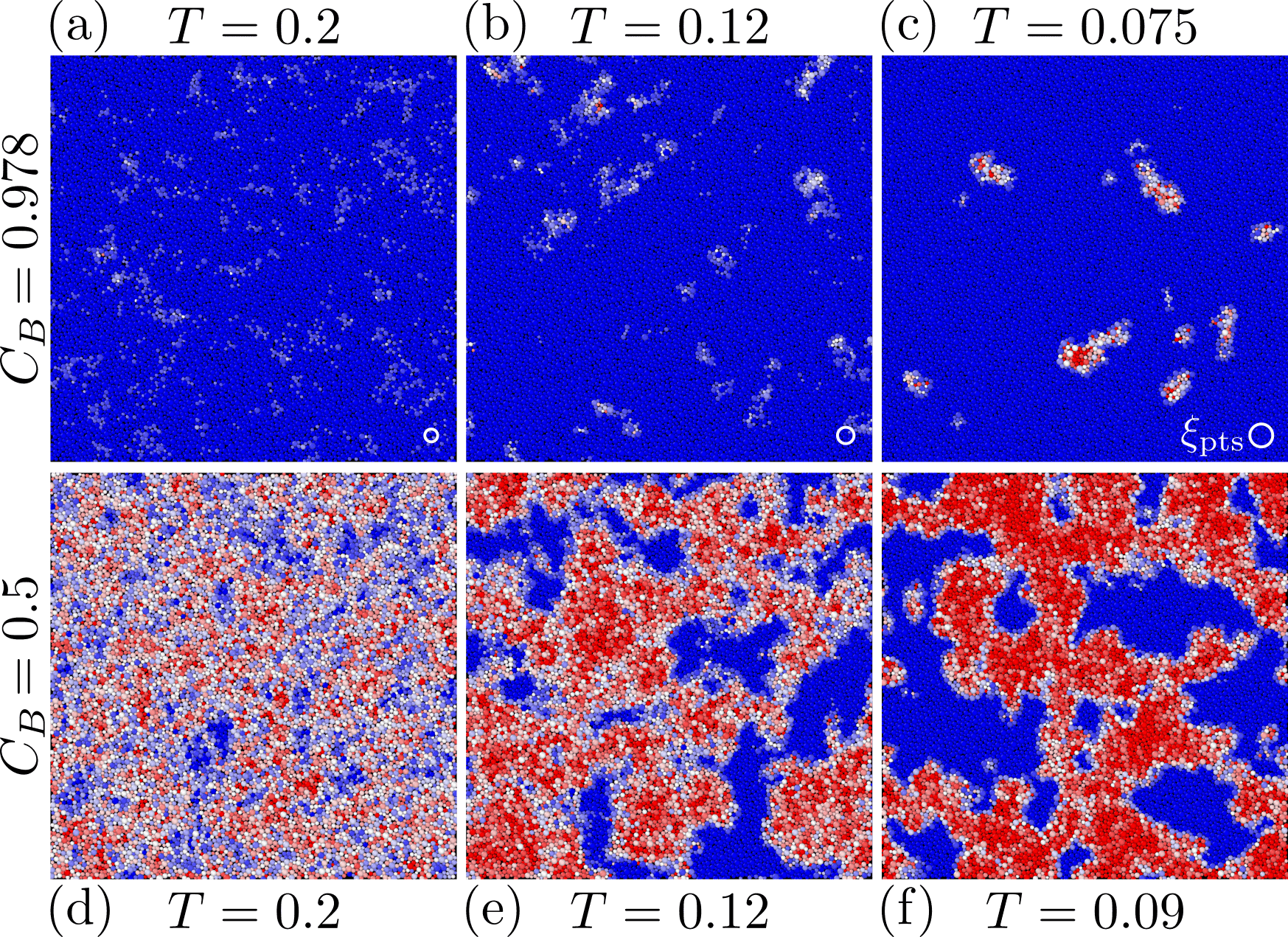}
    \caption{{\bf Temperature evolution of dynamic heterogeneity in $2d$.} 
      Particles are coloured following the bond-breaking correlation $C_B^i$, see legend in Fig.~\ref{fig:redblue}. (Top, a-c) Relaxation at early times $t \ll \ta$ where $C_B(t) = 0.978$, at different temperatures. Snapshots are taken at $t = 6, 2\times 10^3, 3.2 \times 10^8$ from (a) to (c). The circles indicate the point-to-set lengthscale $\xi_{\rm pts}$ discussed in Fig.~\ref{fig:divergence}.
      (Bottom, d-f) Relaxation close to the $\alpha$-relaxation, where $C_B(t) = 0.5$. Snapshots are taken at $t = 500, 1.5\times 10^5, 1.7\times10^8$ from (d) to (f). The linear size of the box is $L=100$.}
    \label{fig:tempevolution2d}
\end{figure}

We first investigate the very initial stages of the relaxation process, at $t \ll \ta$. We show in Fig.~\ref{fig:tempevolution2d} three snapshots of $2d$ liquids at a time $t$ where the average correlation is $C_B(t) = 0.978$. The temperatures are equal, or close, to three relevant temperatures: $T_o=0.2$ (left), $T_{\rm mct}=0.12$ (middle), and $T = 0.075= 1.07\,T_g$ (right). Clearly, the picture changes with temperature. At the onset temperature $T_o$, many particles turn to have lost a few neighbours, and appear in light blue. These particles are scattered throughout the sample, and do not form well-defined clusters. Moving to the mode-coupling crossover temperature $T_{\rm mct}$, we distinguish relatively extended regions containing only quiescent particles. The particles which have undergone some rearrangements start to cluster. Still, the clusters remain rough, numerous, with ill-defined boundaries. At the lowest temperature $T= 0.075$ closest to $T_g=0.07$, the early stage of relaxation looks qualitatively different. We observe a small number of very mobile (red) particles localised into compact clusters with clear boundaries. These particles have undergone several relaxation events leading to $C_B^i=0$. The particles in the rest of the sample are all quiescent and have not undergone a single rearrangement. 

\begin{figure}
    \includegraphics[width=\columnwidth]{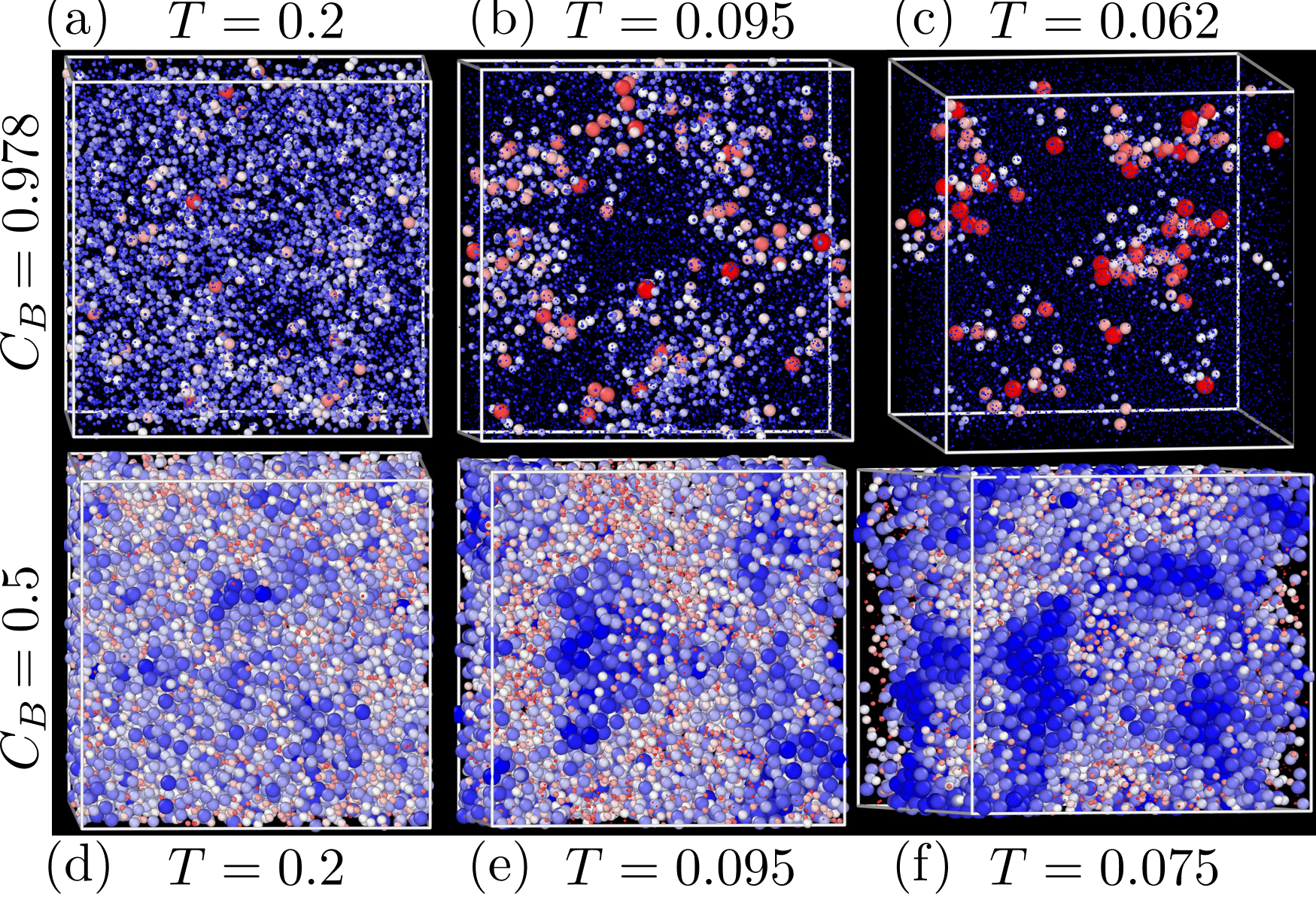}
    \caption{{\bf Temperature evolution of dynamic heterogeneity in $3d$.}
    (Top, a-c) Relaxation at early times $t \ll \ta$ where $C_B(t) = 0.978$, at different temperatures taken at $t=6.5$, 475, and $1.8 \times 10^7$ from (a) to (c). (Bottom, d-f) Relaxation close to the $\alpha$-relaxation, where $C_B(t) = 0.5$ at different temperatures taken at $t=45$, $2.7 \times 10^4$, $1.2 \times 10^7$ from (d) to (f). The particles are coloured following their bond-breaking correlation (see Fig.~\ref{fig:redblue}), and drawn with a diameter proportional to either $1-C_B^i(t)$ (top) or $C_B^i(t)$ (bottom). The linear size of the box is $L=21.5$.}
    \label{fig:tempevolution3d}
\end{figure}

Turning to larger times where structural relaxation occurs, $C_B(t)=0.5$, we again observe a dramatic temperature dependence. The dynamics is almost spatially homogeneous at the onset temperature and becomes spatially correlated near $T_{\rm mct}$ where relatively well-defined domains have emerged, yet with large fluctuations within them and rough boundaries. As the temperature is lowered further, we observe again a much clearer contrast between relaxed and unrelaxed domains which become more compact and less irregular at their boundaries. \rev{Therefore, even though the average correlation $C_B(t)$ is the same within the top and bottom panels of Fig.~\ref{fig:tempevolution2d}, the distribution of local values $C_B^i(t)$ evolves from a single peak around the average at high temperatures to a nearly bimodal distribution, peaking at 0 and 1 at low temperature.} We also note that the characteristic size of the heterogeneity shown in these plots changes a lot between $T_o$ and $T_{\rm mct}$, but appears to evolve more slowly below $T_{\rm mct}$. We shall also document this point below, see Fig.~\ref{fig:divergence}.

In Fig.~\ref{fig:tempevolution3d} we show that similar conclusions are supported by our simulations in $3d$ although snapshot rendering is more intricate. To ease visualisation, not only particles colours but also diameters follow the bond-breaking correlation $C_B^i$: at short (resp. large) times, mobile (resp. quiescent) particles are shown with a larger diameter. In $3d$, the emergence at very low temperatures of well-defined mobile clustered at short times is very clear, whereas a very diffuse pattern is observed at higher temperatures. At large times and low temperature, extended clusters of quiescent particles (blue) are visible. We instead notice that it is difficult to observe the dynamic heterogeneity field near the mode-coupling crossover in $3d$. 
\section{Early times of the structural relaxation}

\label{sec:clusters}

\subsection{Van Hove distribution functions}

\begin{figure}
  \includegraphics[width=\columnwidth]{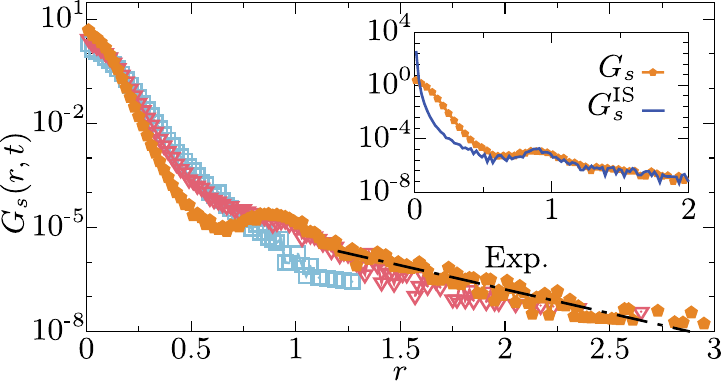}
\caption{{\bf Heterogeneous dynamics at early times.} Van Hove distribution of particle displacements $G_s(r,t)$ in $3d$ at the time when $\Delta^{\rm IS}(t)=6 \times 10^{-3}$ (dash-dotted line in Fig.~\ref{fig:disp_3d}) for $T=0.095$, 0.08 and 0.059 [same legend as in Fig.~\ref{fig:timecorrel}(a)]. The dash-dotted line is an exponential fit of the tail of the distribution at the lowest temperature. The inset compares the van Hove distributions computed at $T=0.059$ either from real displacements or from the visited inherent states (IS).}
\label{fig:vanhove}
\end{figure}

To analyse particle motion at very short times, we first record the self-part of the van Hove distribution function
\begin{equation}
    G_s(r,t) = \frac{1}{4\pi r^2}\langle \delta ( r - |{\bm r}_i(t) - {\bm r}_i(0) ) |\rangle,
\end{equation}
in the three-dimensional liquid, see Fig.~\ref{fig:vanhove}. To follow the evolution with temperature of this quantity we select times where the mean-squared displacement in the inherent structure is constant, $\Delta^{\rm IS}(t)=6 \times 10^{-3}$, which is indicated as a horizontal line in Fig.~\ref{fig:disp_3d}. This threshold corresponds to the value of $\Delta^{\rm IS}$ as the MSD $\Delta$ departs from its plateau. As such, we probe the distribution of particle displacements at extremely early times compared to $\tau_\a$. 

At all temperatures, the distribution displays a peak around $r=0$, caused by particles vibrating around their position at $t=0$. Although thermal vibrations occur on a lengthscale $r \approx 0.15$, the van Hove distribution extends to much broader values, with a tail which is well-described by an exponential decay~\cite{chaudhuri2007universal}. The exponential tail encompasses displacements that exceed the particle size, and extends to an increasing range of particle displacements as temperature lowers. When $T \leq 0.07$, a peak emerges near $r \approx 0.9$, which coincides with the position of the first peak of the pair distribution function $g(r)$. Therefore, the emerging peak in $G_s(r,t)$ is due to particles hopping to the position previously occupied by one of their neighbours. 

This behaviour means that dynamic heterogeneity becomes more pronounced as $T$ decreases, even at $t \ll \tau_\a$. On these short timescales, a large majority of particles is quiescent. These particles have very small displacements due to thermal vibrations, and they populate the broad peak near $r=0$ in $G_s(r,t)$. However, a very small fraction of particles is already quite mobile and displays displacements that are broadly distributed and reach several particle diameters long. These mobile particles give rise to the exponential tail in $G_s(r,t)$. In real space, they correspond to the rare clusters of relaxed particles detected at short times in Fig.~\ref{fig:redblue}. This small population of mobile particles is responsible for the slow growth of the MSD computed within inherent states in Fig.~\ref{fig:disp_3d}. This is confirmed in the inset of Fig.~\ref{fig:vanhove} which demonstrates that the exponential tail is unaffected by removing thermal fluctuations. Finally, the temperature evolution of the van Hove distribution also confirms that the contrast between mobile and quiescent particles becomes stronger at lower temperatures, which leads to the stronger contrast observed in the relaxation snapshots in Sec.~\ref{sec:visualisation}.

\subsection{Complex particle motion inside isolated clusters}

The van Hove distribution at short times and low temperatures shows that a small population of highly mobile particles coexists with a majority of particles which simply undergo thermal vibrations. Yet, the distribution \rev{itself} provides no information as to how these mobile particles are organised in space and what type of rearrangements they have undergone.

\begin{figure}
\includegraphics[width=\columnwidth]{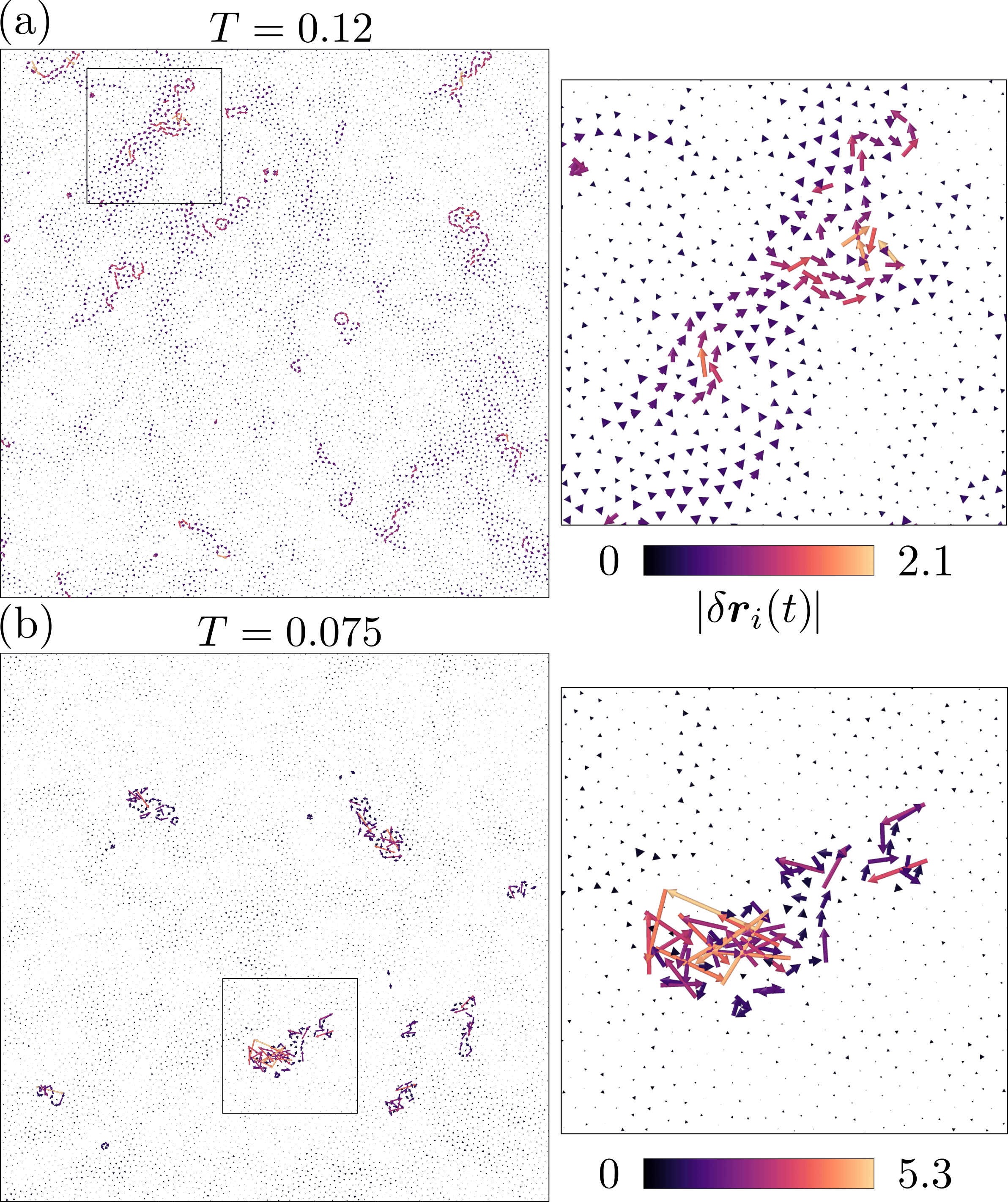}
\caption{{\bf Displacement field at early times.}
Displacement fields at (a) $T = T_{\rm mct} = 0.12$ and (b) $T= 0.075$, corresponding to the images in Fig.~\ref{fig:tempevolution2d}(b,c). Arrows show the particle displacement vectors $\delta \bm{r}_i(t) = \bm{r}_i(t)-\bm{r}_i(0)$ at (a) $t= 2\times 10^3$, (b) $3.2\times 10^8$. The linear size of the box is $L=100$. \rev{Right: zoom on the squared region, displacement vectors are coloured by their magnitude.}}
\label{fig:smallclusters}
\end{figure}

In Fig.~\ref{fig:smallclusters} we highlight these mobile particles and their displacements for \rev{$T=T_{\rm mct}$, and for $T=0.075$} near $T_g$ in the two-dimensional system. We represent with arrows the displacement vectors of the particles $\delta \bm{r}_i(t) = \bm{r}_i(t)-\bm{r}_i(0)$, at $t=2\times 10^3$, $3.2\times 10^8$ for (a) and (b), respectively. These two displacement fields correspond to the images in Fig.~\ref{fig:tempevolution2d}(b,c). \rev{The vectors colour codes for their magnitude, from dark to bright,} with a maximum magnitude clearly depending on temperature. \rev{We observe a clear spatial heterogeneity in particle displacements, with a coexistence of mobile particles characterised by large displacement vectors and
quiescent particles with very small displacements. The displacement field in the quiescent regions qualitatively resembles the structure of the long-wavelength modes among the eigenstates of the Hessian matrix of amorphous glassy states~\cite{widmer2008irreversible,hocky2013small,brito2009geometric}.  }


\rev{Yet, a careful look reveals that the physics is qualitatively different at the two temperatures. These differences are most appreciated on the right panels, where we magnify some mobile regions. Near $T_{\rm mct}$, the mobile cluster contains particles which have moved at most $|\delta \bm{r}_i(t)| \approx 1$. The displacements are coherent, extending over distances up to ten particles. This explains the partial decorrelation of the structure, as shown in Fig.~\ref{fig:tempevolution2d}(b). These displacements resemble string-like motion~\cite{donati1998stringlike}.}


\rev{In comparison, mobile clusters close to $T_g$ are composed of particles that have moved significantly larger distances and have fully decorrelated the initial structure. We still observe mode-like displacements in the unrelaxed regions, although with a much smaller amplitude. The mobile particles are clearly identified, and they form clusters of various sizes and geometries, which are quite compact with sharp boundaries. The inset reveals that mobile particles have undergone a large number of rearrangements, leading to displacement vectors that are very entangled.}

\rev{At very low temperatures, the physical picture is that of highly localised rearrangements taking place in an otherwise elastic matrix responding as an amorphous solid. It would be interesting to measure spatial correlations of these small displacements for several temperatures, as they could indicate a possible crossover from delocalised mode-like motion at high temperature to elastic-like displacements at lower temperatures.}

Inside the clusters of mobile particles, the particle displacements appear complicated and do not form simple patterns such as strings~\cite{donati1998stringlike,yu2017structural}, loops or simple swaps. Anticipating on Sec.~\ref{sec:towards}, we understand that these complex displacement patterns result in fact from the accumulation of a large number of elementary relaxation events which take place inside these localised clusters.    

\subsection{Cluster analysis: statistical properties}

\label{sec:cluster}

One way to quantify the statistical properties of the localised clusters where structural relaxation occurs at early times is to perform a cluster analysis. We first threshold particle mobility into two families, and then consider that two mobile particles belong to the same cluster if their relative distance is smaller than $1.5$, corresponding to the first minimum in the total pair distribution function. With these definitions, we can group relaxed particles into independent clusters and perform an analysis of the statistical properties of the clusters observed in the simulations. A similar analysis has been performed before in the regime above $T_{\rm mct}$~\cite{kob1997dynamical,donati1998stringlike,starr2013relationship,chaudhuri2008tracking}. Back then, the procedure required relatively arbitrary thresholding in particle mobility, which instead becomes more physical at the low temperatures investigated here. \rev{We have implemented several thresholding procedures, which all yield the same qualitative picture.}

\begin{figure}
\includegraphics[width=\columnwidth]{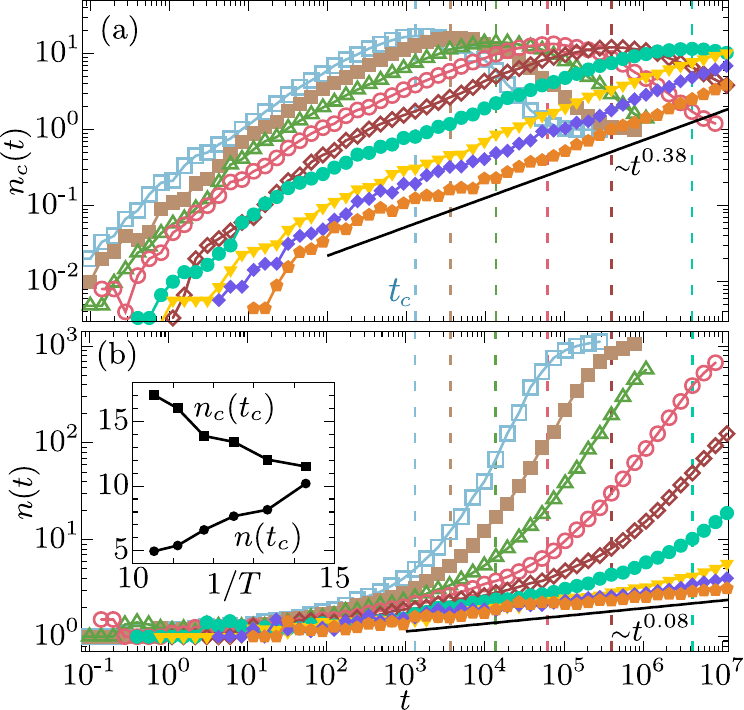}
  \caption{{\bf Statistical analysis of relaxed clusters in $3d$.}
    (a) Average number of clusters $n_c(t)$ of mobile particles for temperatures as in Fig.~\ref{fig:timecorrel}(a). Dashed lines indicate the time $t_c$ at which $n_c$ is maximum. At low temperature, the number of clusters grows as a power-law (line), with exponent $0.38$. (b) Average number of mobile particles $n(t)$ per cluster, with a slow power law growth indicated as a guide to the eye. Inset: average number $n_c(t)$ and size $n(t)$ of clusters at $t=t_c$ as a function of the inverse temperature.} 
\label{fig:clust_3d}
\end{figure}

In Fig.~\ref{fig:clust_3d}(a), we show the average number of distinct clusters $n_c(t)$ as a function of time for different temperatures in $3d$. This quantity is extensive \rev{at short times}, \rev{but} its time and temperature evolution \rev{should not} depend on $N$. At all temperatures, $n_c$ increases with time before eventually reaching a maximum and decreasing at longer times. We have computed the pair distribution function of clusters (not reported), which does not show any peculiar spatial dependence at short times. This suggests that the clusters are, at least initially, randomly distributed in space and uncorrelated. The time dependence of $n_c(t)$ is interesting as the growth towards its maximum is well described by a power law which extends over a broader range as $T$ decreases. The corresponding exponent slowly decreases with decreasing $T$, from $n_c(t) \sim t^{0.5}$ at $T_{\rm mct}$ to $n_c(t) \sim t^{0.38}$ close to $T_g$.

We display in Fig.~\ref{fig:clust_3d}(b) the average number of particles $n(t)$ per mobile cluster. At early times, when the number of clusters starts to grow, their average size $n$ remains modest (about $2$-$3$ particles), but the cluster size is distributed (see below). This analysis confirms the intuition obtained from Fig.~\ref{fig:redblue}: at early times, the dynamics is very heterogeneous in space and time, and structural relaxation is initiated in randomly distributed localised clusters containing only a few particles, which relax in an extended `sea' of blue, quiescent particles. As time increases, the typical size of these relaxed clusters increases and mobile regions slowly coarsen with time. At the lower temperature studied, the growth of $n(t)$ at times much shorter than $\tau_\alpha$ is very slow, as revealed by the power law $n(t) \sim t^{0.08}$ indicated in Fig.~\ref{fig:clust_3d}(b). 

At later times, the number $n_c$ of mobile clusters reaches a maximum at a time $t=t_c$ before decreasing towards $1$ at very long times, when all particles are mobile and the system forms a single mobile cluster~\cite{yu2017structural}. Therefore, we can generically expect the emergence of a maximum number of clusters. At $t_c$, independent clusters start to merge either because two clusters appear at nearby locations, or because two nearby growing clusters merge. Either way, the emergence and growth of mobile clusters should naturally induce a form of mobility percolation as time increases~\cite{yu2017structural}. 

The timescale $t_c$ where $n_c$ is maximal grows with decreasing temperature. From $T = 0.097$ to $0.0715$, the ratio $t_c/\tau_\a$ decreases from $0.3$ to $0.02$, while $\tau_\a$ grows by about $4$ orders of magnitude. Thus, $t_c$ is mostly driven by the evolution of $\tau_\a$ but both timescales are not precisely proportional. The time $t_c$ results from the competition between the emergence of new relaxed clusters and the merging of pre-existing ones. As such, we see no reason to interpret $t_c$ as an important timescale to describe the structural relaxation, in contrast to Ref.~\cite{yu2017structural}. In particular, $t_c$ does not lead to any observable signature in time correlation functions or relaxation spectra.

Our interpretation for the physical origin of a maximum $n_c$ is supported by the steep evolution observed at $t>t_c$ for the average cluster size $n(t)$. Finally the distribution of cluster sizes $P(n,t)$ at $t=t_c$ is also consistent with the physics of percolation, as it shows a power-law decay with a temperature-independent exponent, $P(n,t_c) \sim n^{-1.8}$ (data not shown). Similar distributions have been obtained before~\cite{starr2013relationship,donati1998stringlike}. \rev{The absence of temperature evolution suggests} that the strong thresholding performed at relatively high temperatures seriously weakens the physical evolution observed in Fig.~\ref{fig:redblue}, and may have artificially rendered the high temperature regime more \rev{heterogeneous and clustered} than it really is.

Two characteristic quantities exhibit an interesting temperature evolution, as shown in the inset of Fig.~\ref{fig:clust_3d}. As temperature decreases, the average number of clusters $n_c(t_c)$ decreases, while their typical size $n(t_c)$ increases. This echoes the observation made from Fig.~\ref{fig:tempevolution2d}(d-f) close to the structural relaxation time: relaxed domains become larger and more compact as temperature decreases. The compactification of the domains at the moment they percolate is confirmed by the temperature evolution of the product $n(t_c) \times n_c(t_c)$ which increases by about 50\% (from 80 to 120) in the temperature regime shown in Fig.~\ref{fig:clust_3d}.  

\subsection{Waiting time distribution of cluster relaxation: Emergence of a power law tail}

\label{sec:waiting}

\begin{figure}
\includegraphics[width=\columnwidth]{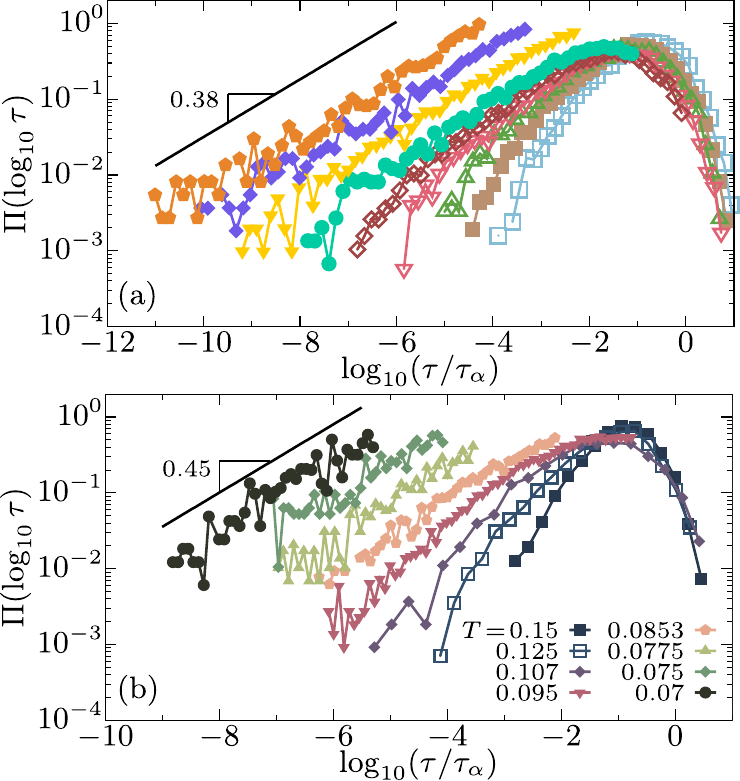}
\caption{{\bf Emergence of a power-law tail in the waiting time distribution of relaxed clusters.} Waiting time distribution $\Pi(\log\tau)$ for the appearance of new clusters of relaxed particles in (a) $3d$ and (b) $2d$. Close to $T_g$, the distribution is well fitted at short times $\tau/\ta\ll 1$ by a power law with an exponent $b \approx 0.38$ (in $3d$) and $0.45$ (in $2d$). Legend as in Fig.~\ref{fig:timecorrel}.}
\label{fig:distribapparition}
\end{figure}

The power-law growth of the number of relaxed clusters $n_c(t)$ suggests that the appearance of a new cluster of mobile particles is a stochastic process. Here we define $\tau$ as the first time $t$ at which a new cluster of mobile particles appears. We have measured the distribution $\Pi(\log\tau)$ of waiting times $\tau$ for new clusters of relaxed particles in both $2d$ and $3d$, see Fig.~\ref{fig:distribapparition}. To avoid statistical noise, we only represent data for temporal bins which have accumulated more than $6$ events among more than $5\times 10^3$ for each temperature. At the lowest temperatures, it is no longer possible to normalise the distributions as we cannot observe the entire range of waiting times for the appearance of relaxed clusters. In the log-log representation of Fig.~\ref{fig:distribapparition} the low-temperature distributions are defined up to an arbitrary vertical shift. Finally, to compare the distributions at different temperatures we normalise the time axis by the average relaxation time $\ta(T)$ at each temperature.  

For temperatures near $T_{\rm mct}$ and slightly below, we measure the entire distribution of waiting times $\Pi(\log \tau)$. Strikingly, even at these relatively high temperatures (from an experimental viewpoint) the distributions are already quite broad since they extend over more than 4 decades of waiting times. Compared to an average relaxation time $\ta \approx 10^4$ around $T_{\rm mct}$, this implies that the width of the distribution is comparable to its average. In other words, there is dynamics happening at all timescales between the microscopic time $t\approx 1\approx \tau_o$ and the structural relaxation time $\ta$.

Exploring the temperature regime below $T_{\rm mct}$ towards $T_g$, we find that this trend becomes even more prominent. The waiting time distributions broaden as $T$ decreases. Interestingly, the functional form of the distribution changes when temperature gets closer to $T_g$ in the sense that it develops a clear power-law tail at $\tau \ll \ta$ 
\begin{equation}
\Pi (\log \tau) \sim \tau^b.
\label{eqn:power_law_distrib}
\end{equation}
The exponent $b(T)$ depends very weakly on temperatures, as it decreases slowly with decreasing $T$. In the vicinity of $T_g$, we measure $b\approx 0.38$ (in $3d$) and $b \approx 0.45$ (in $2d$). This power-law behaviour, which emerges only at $T \ll T_{\rm mct}$ is a novel feature revealed by our numerical strategy which probes equilibrium dynamics near $T_g$ for the first time. 

The similar power-law behaviour found for $\Pi(\log \tau)$ and for $n_c(t)$ in $3d$ should not come as a surprise, as one generally expects that
\begin{equation}
    n_c(t) =\int_{-\infty}^{\log t}\mathrm{d}\log\tau\Pi(\log\tau).
\end{equation}
Plugging Eq.~(\ref{eqn:power_law_distrib}) into the above equation and performing a change of variable directly yields $n_c(t)\sim t^b$, as indeed observed in Fig.~\ref{fig:clust_3d}(a).

\subsection{Short-time dynamics in the frequency domain: Emergence of excess wings}

\begin{figure}
\includegraphics[width=\columnwidth]{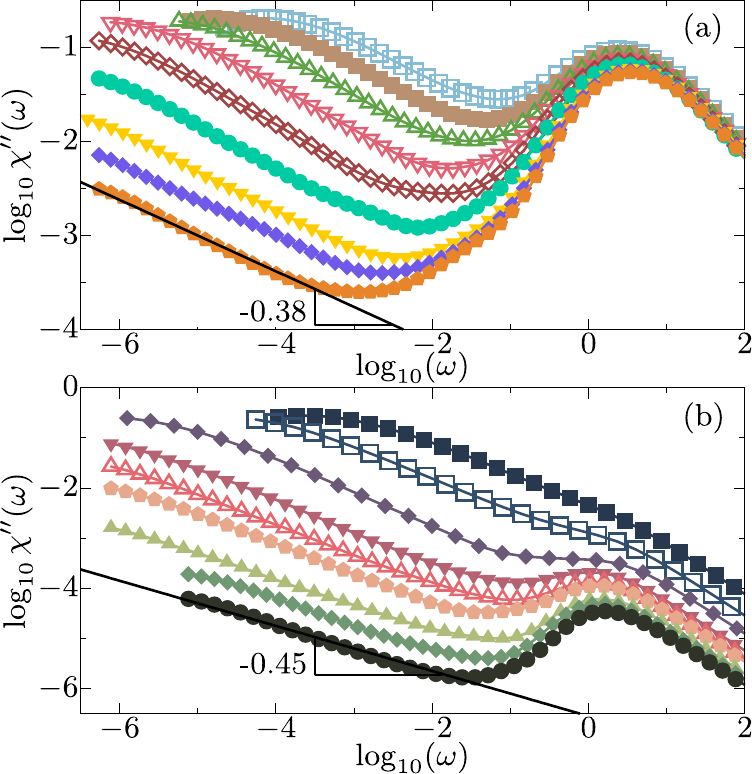}
\caption{{\bf Emergence of excess wings in relaxation spectra near $T_g$.} Equilibrium relaxation spectra $\chi^{\prime\prime}$ as a function of frequency $\omega$ in (a) $3d$ and (b) $2d$. The black lines are power-law fits $\chi^{\prime\prime}(\omega) \sim \omega^{-\sigma}$ with $\sigma \approx 0.38$ (in $3d$) and $\sigma \approx 0.45$ (in $2d$). Legend as in Fig.~\ref{fig:timecorrel}.}
\label{fig:wing}
\end{figure}

In experiments, the relaxation dynamics of supercooled liquids is often probed thanks to spectroscopy techniques (dielectric, mechanical, light scattering, NMR) which measure linear response functions in the frequency domain~\cite{lunkenheimer2000glassy,rossler,flamig2020nmr}. To compare our results with experimental measurements, we define the analog of a susceptibility spectrum in the frequency domain. We assume that the relaxation dynamics stems from a distribution of timescales $G(\log\tau)$, such that a correlation function $C(t)$ which monitors structural relaxation can be written as~\cite{bohmer1998nature}
\begin{equation}
C(t)=\int_{-\infty}^{+\infty} G(\log\tau)e^{-t/\tau}\mathrm{d}\log\tau.
\label{eq:correl_pi}
\end{equation}
This expression amounts to decomposing the broad spectrum of relaxation times characterising supercooled liquids into a series of elementary exponential processes. In the Fourier domain this yields 
\begin{equation}
    \chi(\omega) =  - \int_{-\infty}^{+\infty} 
    G(\log \tau)
    \frac{1}{1+i\omega \tau}\mathrm{d}\log \tau .
\label{eq:chi}
\end{equation}
We defined the relaxation spectrum $\chi^{\prime\prime}(\omega)$ as the imaginary part of the susceptibility $\chi(\omega)$. Following earlier work, we express the distribution of timescales $G$ from the measured time correlation function through the relation $ G \approx -\mathrm{d}C/\mathrm{d}\log t$~\cite{berthier2005numerical}. This allows us to compute numerically an analog of the out of phase susceptibility spectrum $\chi''(\omega)$~\cite{blochowicz2003susceptibility}. In practice, we choose $C(t)$ to be the self-intermediate scattering function in $3d$, and the bond-breaking correlation in $2d$. \rev{While the quantitative detail of the spectra may change for other choices of correlation functions (mainly the relative amplitude of the various processes), the main features are quite robust. This reflects analogous findings of observable dependence in experimental studies~\cite{lunkenheimer2000glassy,rossler,PhysRevE.86.041507,flamig2020nmr}.}  

We report the relaxation spectra in Fig.~\ref{fig:wing} for $d=2,\ 3$. The spectra exhibit a peak at high frequency $\omega \approx 1$, reflecting the rapid decay of the time correlation functions on a microscopic timescale corresponding to thermal motion within a well-defined cage. The lower amplitude of the microscopic peak in $2d$ compared to $3d$ is easily explained. The definition of $C_B$, with a larger cutoff employed to define neighbours at later times, indeed makes it less sensitive than $F_s$ to such motion.

The second peak, found at lower frequency, corresponds to the $\alpha$-relaxation peak and is typically located at frequency $\omega_\a \approx 1/\ta$. At the lowest temperatures, the $\a$-peak exits the numerically accessible frequency window and we are simply left with the high-frequency flank of the structural relaxation peak. 

In both $2d$ and $3d$ the spectra display a power-law behaviour for the lowest temperatures near $T_g$
\begin{equation}
\chi''(\omega) \sim \omega^{-\sigma},
\label{eq:wing}
\end{equation}
where $\sigma(T)$ is an exponent that appears to depend very weakly on temperature and to slowly decrease as $T$ decreases. The power law in Eq.~(\ref{eq:wing}) emerges in the range $\omega\in[10^{-6},\,10^{-3}]$, with an exponent $\sigma \approx 0.38$ in $3d$ and $\sigma  \approx 0.45$ in $2d$. 
The exponents $\sigma$ appearing in the spectra at high frequencies are much smaller than the stretching exponents $\beta=0.56$ (in $3d$) and $0.6$ (in $2d$), \rev{obtained from fitting the corresponding time correlation functions in the $\alpha$-relaxation regime}. Therefore the power law revealed by Eq.~(\ref{eq:wing}) cannot be understood as the high-frequency limit of the $\alpha$-peak. In Ref.~\cite{guiselin2022microscopic} we demonstrated that the power laws correspond to the excess wings appearing in the spectra of many molecular liquids. We emphasise that the excess wings only become clearly visible at temperatures $T\gtrsim T_g$, explaining why they have not been reported before in numerical simulations. We also notice that they do not take the form of a `$\beta$-peak' separated from the $\alpha$-peak, as argued recently in Ref.~\cite{yu2017structural}\rev{, and do not require an internal degree of freedom}. 

Assuming that the distribution of relaxation times $G(\log \tau)$ in Eq.~(\ref{eq:chi}) is dominated by the appearance of new clusters of mobile particles at short times, the power law $\Pi(\log\tau) \sim \tau^{b}$ would translate into $G(\log\tau) \sim \tau^b$. Plugging this expression into Eq.~(\ref{eq:chi}) and performing a change of variable leads to $\chi^{\prime\prime}(\omega)\sim \omega^{-b}$. This suggests that the exponent $\sigma$ describing $\chi''(\omega)$ is equal to the exponent $b$ describing the excess wing. For the two systems studied here, we indeed get a good agreement with $b \approx \sigma \approx 0.38$ in $3d$ and $b \approx \sigma \approx 0.45$ in $2d$. The quantitative agreement between the measured exponent of the excess wings and the waiting time distribution of relaxed cluster shows that the latter process provides a microscopic explanation for this characteristic spectroscopic signature in supercooled liquids near $T_g$. 

\subsection{Suppression of excess wings in ultrastable glasses}

\label{sec:suppression}
\begin{figure}
    \includegraphics[width=\columnwidth]{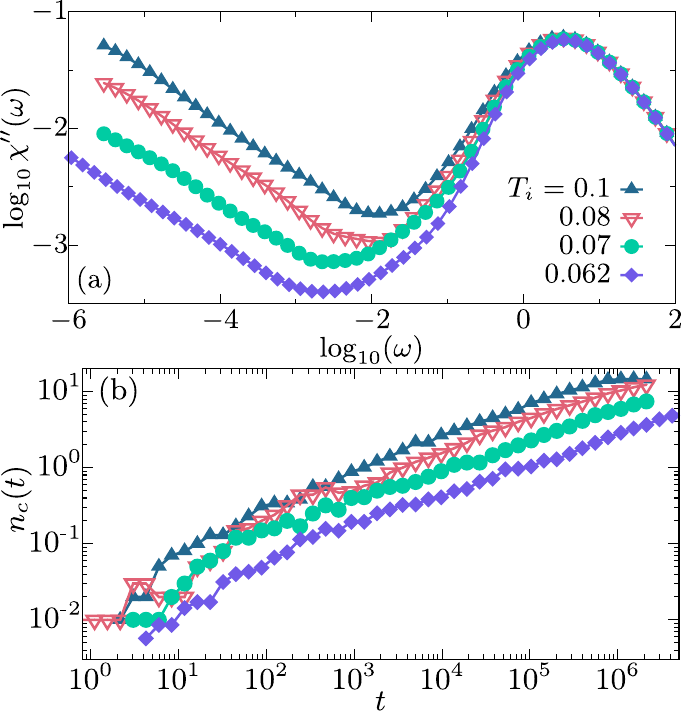}
  \caption{{\bf Suppression of excess wings in ultrastable glasses in $3d$.}
    (a) Spectra $\chi^{\prime\prime}(\omega)$ measured at $T = 0.062$ in glasses first equilibrated at $T_i$, then quenched at $T=0.062$ where they age during $t_w=2 \times 10^6$. We report the equilibrium spectrum, labelled by $T_i = 0.062$. (b) Average number of clusters $n_c(t)$ of mobile particles with the same definition as in Fig.~\ref{fig:clust_3d}.}
    \label{fig:aging}
\end{figure}
To characterise the physical processes responsible for excess wings (or, more generally, of secondary processes) in molecular liquids, relaxation spectra have been measured in non-equilibrium situations~\cite{schneider2000excess,yu2015suppression}. For instance, aging studies report a slow decrease of the amplitude of secondary processes with the waiting time $t_w$ after a rapid quench below $T_g$. This suggests that secondary processes become weaker as the glass is annealed towards more stable states. This effect was directly evidenced in recent experiments using physical vapor deposition as a way to efficiently synthesise glasses with profoundly different degrees of stability~\cite{yu2015suppression}. This would correspond to varying the waiting time in aging experiments, albeit over inaccessible long timescales. The study unequivocally confirmed the suppression of secondary processes in extremely stable glasses. 

To assess whether a similar behaviour is observed numerically, we have investigated the dynamics of glasses with varying stability at temperature $T=0.062$ in $3d$. We vary glass stability by first preparing equilibrium liquids at three temperatures $T_i= 0.07$, 0.08, and 0.10 using the swap Monte Carlo algorithm. Using conventional molecular dynamics with a Nos\'e-Hoover thermostat, we then quench them suddenly to $T=0.062$. The resulting glasses are aged at $T=0.062$ during $t_w=2 \times 10^6$. We have checked that our measurements do not sensitively depend on $t_w$: it is large enough to ensure that all important structural aging processes have shifted outside the observation time window. The temperature $T_i$, which may be interpreted as a fictive temperature~\cite{tool1946relation}, mainly controls the degree of stability of the resulting glasses. A similar strategy was adopted in Refs.~\cite{khomenko2020depletion,Ozawa2018,wang2019low}. Using the relaxation time $\ta(T_i)$ as a quantitative measure of stability, the range of temperatures $T_i$ translates into a variation in glass preparation times between $10^3$ and $10^{12}$, which represents a significant dynamic range.   

In Fig.~\ref{fig:aging}(a), we report the relaxation spectra measured from $F_s(t)$ with conventional MD simulations starting from the samples aged during $t_w$. All spectra exhibit a similar frequency dependence with a power law following Eq.~(\ref{eq:wing}) and an exponent $\sigma$ that appears almost independent of the glass stability. Still, the glass stability strongly affects the amplitude of the relaxation so that, 
\begin{equation}
    \chi''(\omega) \approx A(T_i) \omega^{-\sigma},
    \label{eq:wingaging}
\end{equation}
with a prefactor $A(T_i)$ which decreases by more than one order of magnitude from $T_i = 0.1$ to 0.062. Our measurements directly reveal that secondary processes are suppressed in glasses of increasing stability, in excellent agreement with experimental observations~\cite{yu2015suppression} where a suppression factor of 3 has been achieved. 

Our simulations allow us to uncover the physical origin of this suppression. Following the cluster analysis of Sec.~\ref{sec:clusters}, we present in Fig.~\ref{fig:aging}(b) the time dependence of the number $n_c(t)$ of clusters of mobile particles in aged glasses. All curves display the same time dependence, essentially a power law, with a prefactor that changes by one order of magnitude depending on $T_i$. This observation extends the link established above between the relaxed clusters observed at short times and the excess wing observed in relaxation spectra to non-equilibrium glasses. 
    
\begin{figure}
    \includegraphics[width=\columnwidth]{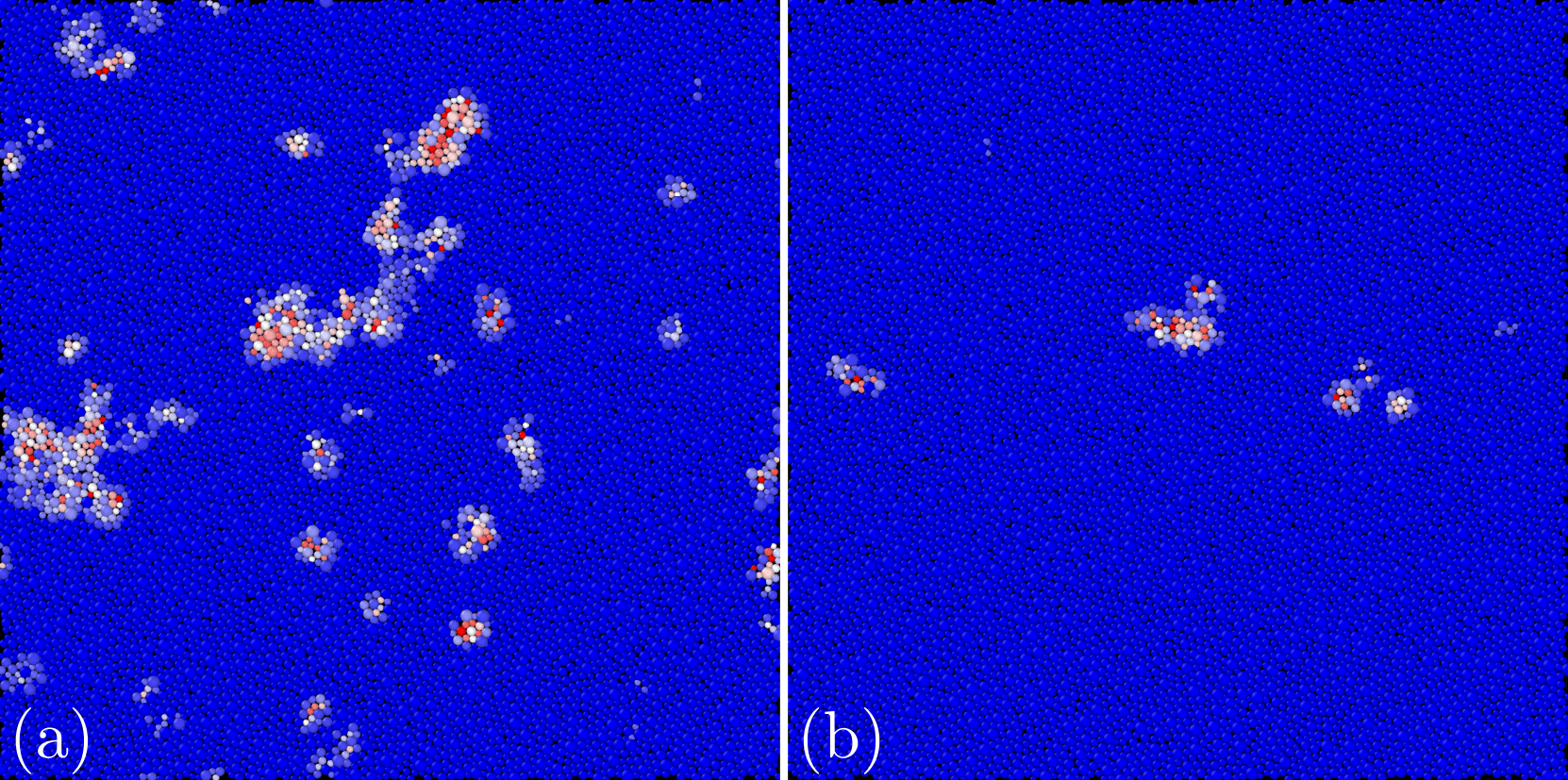}
  \caption{{\bf Relaxed clusters at equal times in poorly aged glasses and equilibrium liquids in $2d$.} Snapshots of two systems simulated for $t=10^7$ at temperature $T=0.075$. Initial conditions are: (a) Sample equilibrated at $T_i=T_{\rm mct}=0.12$, then aged for $t_w=4\times 10^7$ at $T=0.075$, and (b) equilibrium sample at $T_i=0.075$. Particles are coloured following the bond-breaking correlation, as in Fig.~\ref{fig:redblue}. The linear size of the box is $L=100$.}
    \label{fig:agingimage}
\end{figure}     
    
Finally, we illustrate this point thanks to simulation snapshots drawn from $2d$ simulations. We repeat a protocol similar to $3d$, starting from an equilibrium configuration at $T_i=0.12=T_{\rm mct}$, which is suddenly quenched to $T=0.075$ where it ages for a long waiting time $t_w = 4 \times 10^7$ using conventional canonical molecular dynamics. We then run the dynamics over a duration $t=10^7$ and record particle motion. We compare the results to the equilibrium dynamics obtained over the same duration and temperature $T=0.075$. In Fig.~\ref{fig:agingimage} we compare the corresponding mobility fields measured in the poorly annealed system ($T_i=0.12$) and the very stable one ($T_i=0.075$). These images directly confirm that the number and size of the relaxed clusters in the two systems are very different. The depletion of short-time excitations as stability increases explains the suppression of excess wings and of secondary relaxations observed in ultrastable glasses. This finding adds a new item to the growing list of glassy excitations that get depleted as glass stability is varied~\cite{wang2019low, khomenko2020depletion, scalliet2017absence, Scalliet2019, Ozawa2018}. 

The overall physical conclusion is that the population of localised mobile clusters observed at the early times during the equilibrium relaxation of supercooled liquids, becomes stability-dependent when studying the non-equilibrium dynamics of glasses. Many more relaxation events are observed in less stable glasses (a factor of ten in Fig.~\ref{fig:aging}). \rev{The nature of these additional events is different from equilibrium ones,} as they necessarily correspond to irreversible events which slowly drive the system towards equilibrium. Our measurements for the amplitude $A(T_i)$ implies that 90\% of the secondary processes at play in poorly stable glasses prepared using conventional numerical methods are actually not present in equilibrium conditions~\cite{yu2017structural,yu2018fundamental} and \rev{great care must be paid regarding the physical interpretation of such studies.}

\section{How structural relaxation unfolds from early to late times}

\label{sec:towards}

\subsection{Maps of local relaxation time}

\label{sub:direct}

We first illustrate how structural relaxation takes place in space and time starting from the early mobile clusters characterised above. We generate a spatial map of local relaxation times for the long trajectory $t_{\textrm{max}} = 8.2 \times 10^{8}$ at $T=0.09$ of Fig.~\ref{fig:redblue}. Such a representation of the spatially heterogeneous dynamics was used long ago~\cite{foley1993spatial,harrowell1993visualizing,hurley1995kinetic}. To this end, we attach to each particle its relaxation time $\ta^i$ defined as the first time at which particle $i$ becomes mobile, keeping our definition of mobility from the inequality $C_B^i(\ta^i) \leq 0.5$. 

\begin{figure}
\includegraphics[width=\columnwidth]{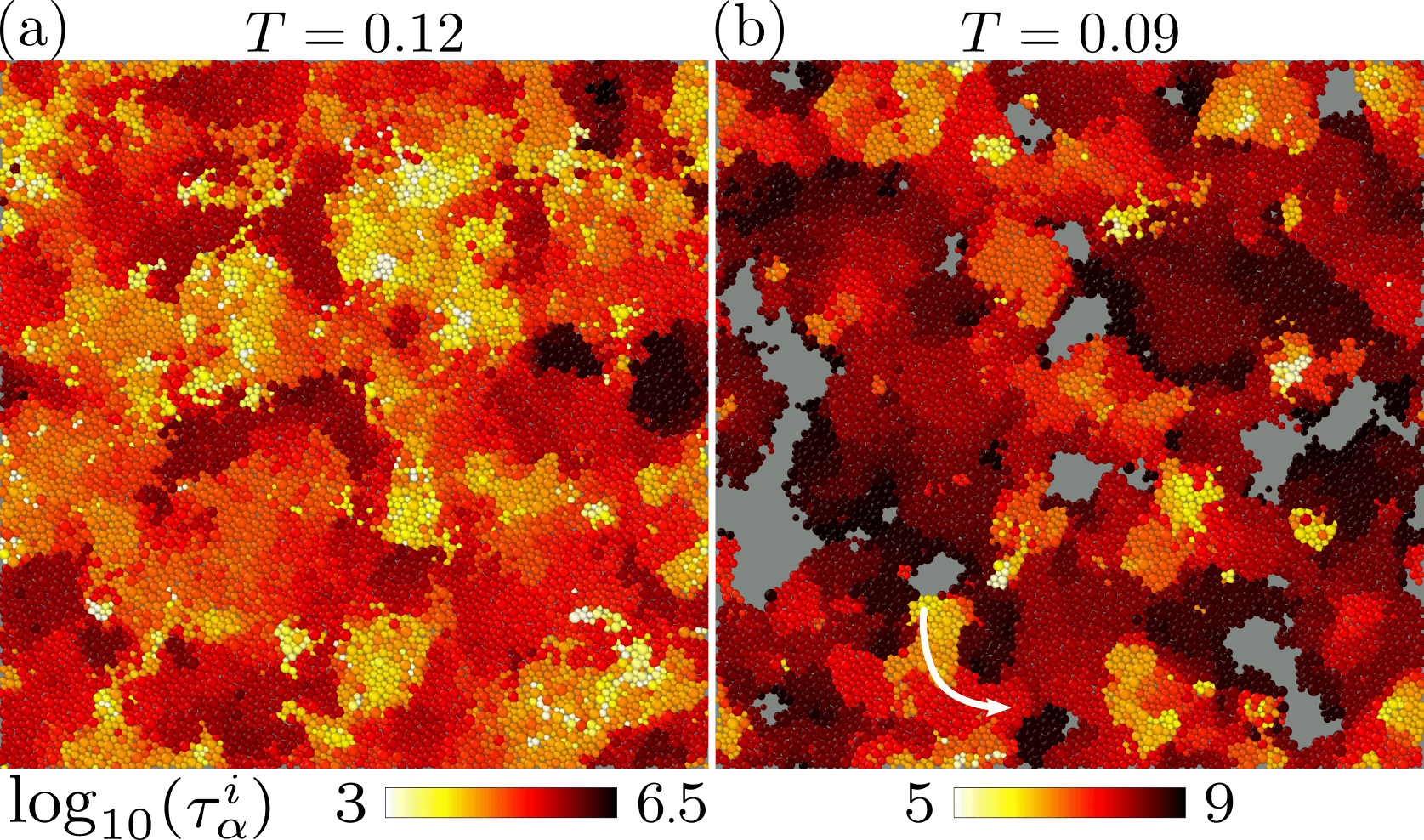}
\caption{{\bf Maps of local relaxation time reveal the emergence of dynamic facilitation.} Map of a local relaxation time $\tau_{\a}^i$ in the two-dimensional system defined as the first time at which $C_B^i \leq 0.5$ for the two-dimensional system and (a) $T=0.12 = T_{\rm mct}$ and (b) $T=0.09$. The colour codes for the value of $\tau_{\a}^i$ are on a logarithmic scale. In (b), the arrow highlights the emergence of dynamic facilitation, and grey regions have not yet relaxed at $t_{\rm max}=8 \times10^8$. The box size is $L=100$.} 
\label{fig:localmap}
\end{figure} 

We show in Fig.~\ref{fig:localmap} two representative maps of the local relaxation time obtained at (a) $T=0.12 \approx T_{\rm mct}$ and (b) a temperature $T=0.09$ halfway between $T_{\rm mct}$ and $T_g$. The latter corresponds to the long trajectory of Fig.~\ref{fig:redblue} which lasts $t_{\rm max} = 8 \times 10^8$. The grey regions in Fig.~\ref{fig:localmap}(b) contain the particles that have not yet relaxed after $t_{\rm max}$. The colour code uses a logarithmic scale from bright/yellow at short times to dark/black at long times. As found in Fig.~\ref{fig:distribapparition}, there are particles which become mobile at extremely short times. Note that the shortest $\ta^i$ are bounded by the time delay $\Delta t$ used to store configurations along the trajectory [we use $\Delta t = 10^3$ and $10^5$ for (a) and (b), respectively]. The upper bound is set by the longest local relaxation time in (a) [about 20$\tau_\alpha$] or by the simulation time $t_{\rm max}$ in (b).

The images in Fig.~\ref{fig:localmap} illustrate the very broad waiting time distributions for the appearance of new mobile particles in the sample. We observe in particular many zones where particles become mobile orders of magnitude before the average structural relaxation time $\ta$. They appear as bright yellow zones, and coexist with regions that instead have a relaxation time larger than $\ta$. We identify many more fast zones near $T_{\rm mct}$ than at $T=0.09$.  
The slow coarsening of relaxed regions observed in Fig.~\ref{fig:redblue} translates into a smooth spatial spreading of the colour in these relaxation time maps. Mobility appearing in a localised region at early time extends at longer times to its neighbouring regions which thus appear darker. We also observe that the typical size of the correlated regions seems to increase on a logarithmic scale: bright yellow domains are very small while darker domains are larger. These images thus confirm that the relaxation starting at localised regions at early times, slowly spreads to neighbouring regions over logarithmically increasing times. 

This very clear spatio-temporal view of the structural relaxation becomes actually much more complicated at higher temperatures. Around the mode-coupling temperature in Fig.~\ref{fig:localmap}(a), the relaxation starts at early times (bright) at numerous locations that are rather fuzzy. Similarly, the spread of mobility to neighbouring regions is much harder to \rev{identify.} This observation suggests that dynamic facilitation is much more clearly defined close to $T_g$ than in the higher temperature regime explored by conventional simulations.

\rev{We also remark that} mobility does not spread isotropically. On the bottom left of Fig.~\ref{fig:localmap}(b) one can indeed distinguish a domain which relaxes very early, surrounded in one direction by a facilitated region identified by a smooth colour gradient to the south (indicated by an arrow), while to the north, one distinguishes a grey region in which no relaxation takes place over the subsequent four decades. Therefore, dynamic facilitation acts isotropically on average, but it appears to be locally anisotropic. This suggests that \rev{the local disorder} plays an important role and controls how mobility can propagate in space. This observation seems in harmony with the anisotropic kinetic constraints introduced in certain kinetically constrained models~\cite{jackle1991hierarchically,garrahan2003coarse,berthier2005numerical}.

\begin{figure}
\includegraphics[width=0.5\columnwidth]{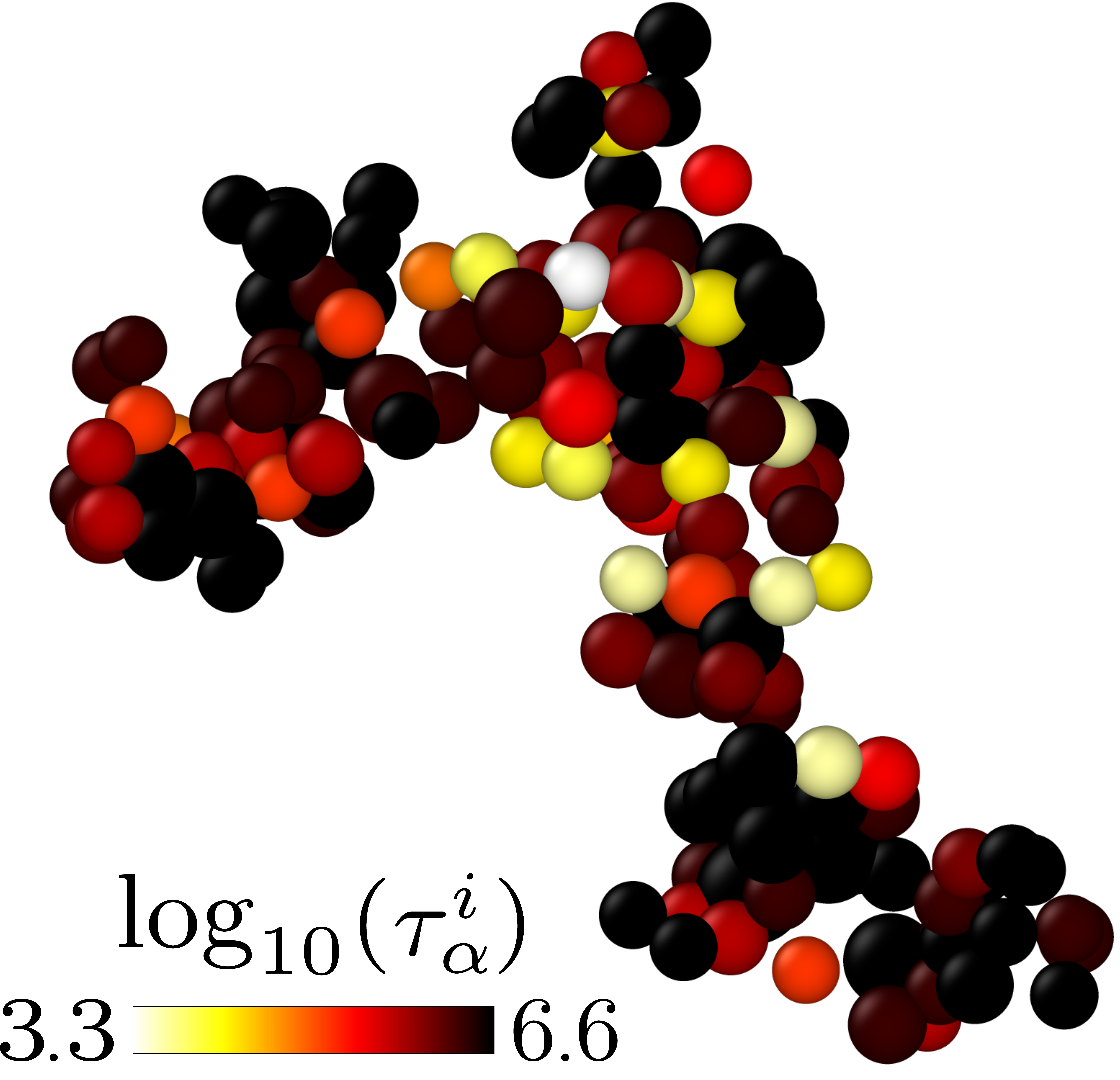}
\caption{{\bf Dynamic facilitation in the $3d$ model.}
  A connected cluster containing 163 mobile particles at $T=0.07$ inside a $3d$ system composed of $N=10^4$ particles (linear size $L = 21.5$) at time $t=4.1\times 10^6$. The particles are represented at their position at $t=0$ using a colour which codes for $\tau_{\a}^i$.}
\label{fig:snap_3d}
\end{figure}

In three-dimensional liquids, we measure the local relaxation time and display how the dynamics evolves in space and time. To ease visualization, we show in Fig.~\ref{fig:snap_3d} a selected relaxed cluster of particles at very low temperature $T=0.07$. Following Fig.~\ref{fig:localmap}, the colour code indicates the local relaxation time. As for $2d$ systems at low temperature, mobility starts at very early times in small localised regions, and spreads to the neighbouring particles over time to form a larger cluster of mobile particles at larger times. Qualitatively, dynamic facilitation thus appears to also play an important role in $3d$. 

\subsection{Space-time trajectories}

In constructing maps of the local relaxation time, each particle is shown only once with an indication about the first time it becomes mobile. These maps say nothing about what happens after this first relaxation event. To address this point, we construct a $d+1$ space-time trajectory representing the evolution of the mobility field along a time axis for $d=2$ liquids~\cite{chandler2010dynamics}. To define mobility on a given time interval of duration $\Delta t$ we use the usual criterion $C_B^i(t+\Delta t, t) < 0.5$. By stacking the mobility field over consecutive time slices of duration $\Delta t$ we can visualise how mobility propagates in space and time. 

\begin{figure}
\includegraphics[width=\columnwidth]{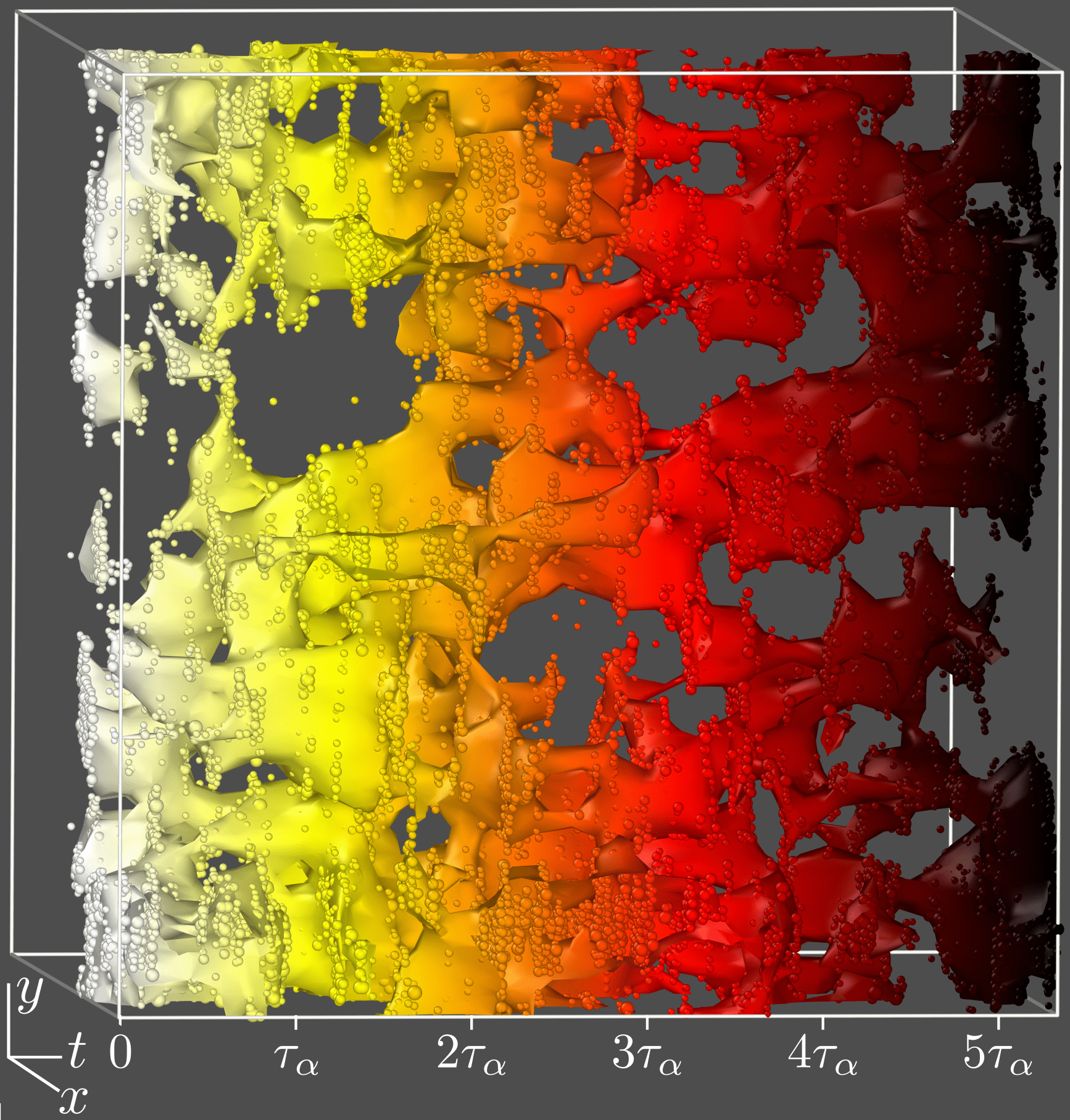}
\caption{{\bf Space-time trajectory in the $2d$ model.} A trajectory starting at $t=0$ (left) finishing at $t_{\rm max}=8\times10^8$ (right) at $T=0.09$ is split into 400 time slices of duration $\Delta t = 2\times10^6$. We show all particles that are mobile within a given time slice, together with a surface representation which draws a contour of the coarse-grained mobility field. To ease $d+1$ visualisation, the colour linearly codes for the time dimension and one half of the system in the $x$-direction is removed.}
\label{fig:bubbles}
\end{figure}

In Fig.~\ref{fig:bubbles}, we show a representative space-time mobility trajectory at $T=0.09$, obtained from the same data as in Fig.~\ref{fig:redblue} and Fig.~\ref{fig:localmap}(b). The trajectory of duration $t_{\rm max} = 8 \times 10^8$ is split into 400 time slices of width $\Delta t = 2 \times 10^6$. The choice of $\Delta t$ results from a compromise. It must be small enough compared to $\ta$ so that we can resolve how the structural relaxation unfolds from early times, but not too small so the fraction of mobile particles within each time interval is non-vanishing. With our choice, about 5\% of the particles are mobile in each slice. Only particles that are mobile with a time slice are rendered. To ease visualisation we only show particles with $0 < x < L/2$. We highlight the main features of the space-time trajectory by constructing a surface representation of the set of mobile particles, as implemented in the Ovito software~\cite{ovito}. 

The most striking novel observation deduced from Fig.~\ref{fig:bubbles} is the emergence of `tubes' in space-time. Physically, this corresponds to localised regions in space which keep relaxing in many successive time slices. We conclude that a large number of relaxation events accumulates in some regions, \rev{even long after} the first relaxation event. In Sec.~\ref{sub:predictability} we quantify this accumulation effect by introducing an appropriate statistical tool. 

In addition, a careful examination of Fig.~\ref{fig:bubbles} reveals that these tubes are not perfectly aligned along the time direction. They can bend, widen, or merge with neighbouring tubes. In terms of particle motion, this means that the relaxation events which accumulate in a localised region are not simply the repetition of the exact same event (for instance a complex but reversible particle motion), which means that relaxation can spread to neighbouring particles. As a result, mobility propagates in space from one time slice to another. This observation accounts for the slow coarsening of the relaxed regions observed in Fig.~\ref{fig:redblue}. In Sec.~\ref{sec:coarsening} below, we quantify this coarsening process. \rev{The propagation of mobility from one region to a neighbouring one via the accumulation of complex localised relaxation events provides the microscopic origin of dynamic facilitation.}

\subsection{Accumulation of localised relaxation events}

\label{sub:predictability}

\begin{figure}
    \includegraphics[width=\columnwidth]{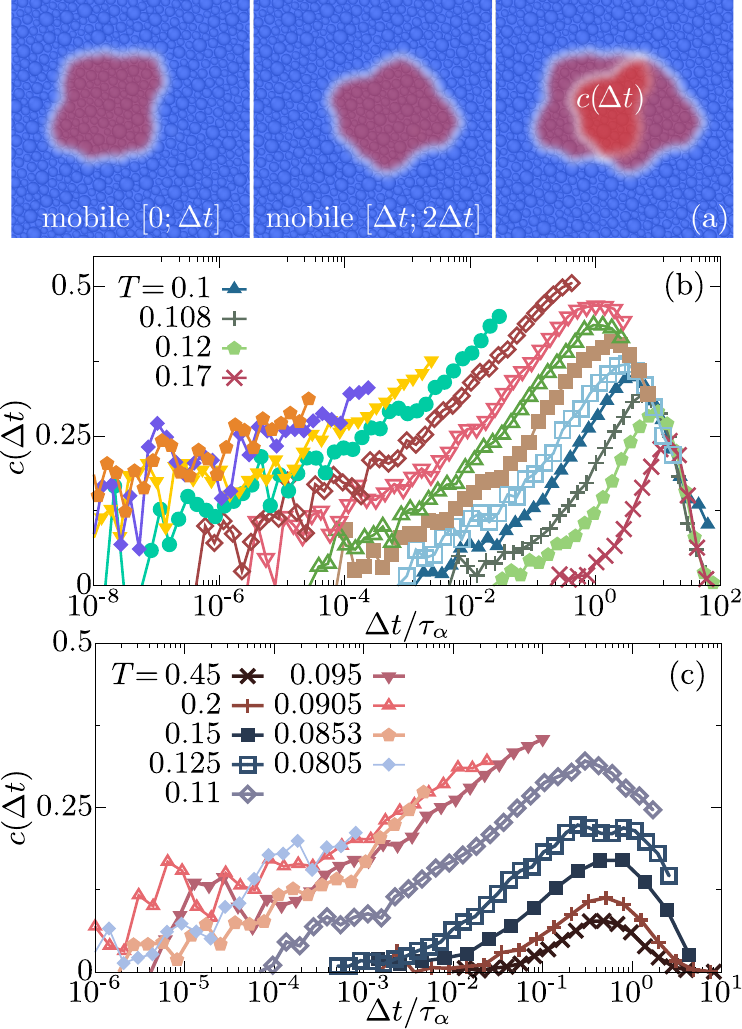}
    \caption{{\bf The accumulation of relaxation events becomes more probable towards $T_g$.} (a) The quantity $c(\Delta t)$ in Eq.~(\ref{eq:c_Delta_t}) quantifies \rev{the probability for particles to be mobile within two consecutive} intervals of duration $\Delta t$. Data for $c(\Delta t)$ in $3d$ [(b), legend provided in Fig.~\ref{fig:timecorrel}(a)] and $2d$ (c) liquids at various temperatures. The time axis is rescaled by $\tau_\a$ measured from $F_s(t)$ and $C_B(t)$, respectively. }
        \label{fig:c_Delta_t}
\end{figure}

To quantify the accumulation of mobility in the same regions over time, we \rev{denote as} $m_1(\Delta t)$ the fraction of particles that are mobile (defined again via $C_B^i$) in the time slice $[t,\, t+ \Delta t]$. We then identify the particles that are mobile over the next time slice $[t+ \Delta t,\, t+ 2\Delta t]$. Finally, we \rev{define $m_2(\Delta t)$ as} the fraction of particles that are mobile in two consecutive time intervals, i.e. mobile in both $[t,\, t + \Delta t]$ and $[t + \Delta t,\, t+ 2\Delta t]$. 

By construction, $m_2 \leq m_1$, the equality being obtained when the dynamics repeats itself exactly from one time slice to the next. In the opposite case of completely uncorrelated dynamics between two consecutive intervals, one gets $m_2 = m_1^2 < m_1$. To quantify the mobility correlation between consecutive time slices, we introduce the quantity $c(\Delta t)$
\begin{equation}
c(\Delta t)=\frac{m_2 - m_1^2}{m_1 - m_1^2},
\label{eq:c_Delta_t}
\end{equation}
whose normalization ensures that $0 \leq c(\Delta t) \leq 1$. By definition, we have $c(\Delta t)=1$ if the dynamics is perfectly correlated over two consecutive intervals, while $c(\Delta t)=0$ if it is \rev{completely} uncorrelated. For a given $\Delta t$, a large value $c(\Delta t)$ implies that a large fraction of mobile particles are identical in successive time intervals of duration $\Delta t$. This is illustrated in Fig.~\ref{fig:c_Delta_t}(a).

We have investigated the evolution of $c(\Delta t)$ as a function of $\Delta t$ and temperature $T$ in both $2d$ and $3d$, see Fig.~\ref{fig:c_Delta_t}. To compare different temperatures, we rescale $\Delta t$ by the average relaxation time $\ta$. The latter is measured from $F_s(t)$ in $3d$ and $C_B(t)$ in $2d$. All curves exhibit a qualitatively similar time dependence with $c(\Delta t) \approx 0$ at both microscopic times $\Delta t \approx 1$ and at very long times $\Delta t \gg \ta$, where the dynamics in successive frames is obviously uncorrelated. The probability $c(\Delta t)$ thus exhibits a maximum at a time which scales almost like $\ta$. The value at the maximum increases with decreasing the temperature. We also see that at any fixed value of $\Delta t / \ta$, the value of $c(\Delta t)$ grows when $T$ decreases. 

These measurements quantitatively confirm the `tube' interpretation of the dynamics detected in the space-time trajectory of Fig.~\ref{fig:bubbles}. At low temperature, the mobility field measured in a given time frame becomes increasingly correlated with the mobility field measured in the consecutive time frame: relaxation events accumulate at identical locations over long times to form tubes in $d+1$ dimensions.  

The existence of a growing maximum at $\Delta t \approx \ta$ shows that the spatial structure of dynamic heterogeneities characterising the structural relaxation becomes increasingly similar from one slice of duration $\ta$ to the next as temperature decreases. For example, at $T=0.075$ in $3d$, we find that $c(\tau_\alpha) \approx 0.5$, which implies that half of the particles that relax over one relaxation time $\ta$ also relax over the next relaxation time. One can physically anticipate that such a mechanism leads to an increased lifetime of dynamic heterogeneities at low temperature. 

\subsection{Slow temperature-dependent coarsening of relaxed domains}

\label{sec:coarsening}

We now investigate how mobility spreads in space over larger timescales. Natural quantities to measure the growth of spatially correlated regions are four-point spatial correlation functions of the mobility field, either in real or in Fourier space~\cite{lavcevic2003spatially,bennemann1999growing,donati1999growing}. Both functions are related to the four-point dynamic susceptibility that has been the subject of a large number of studies~\cite{glotzer2000time,donati2002theory,toninelli2005dynamical,berthier2011dynamical}. It is however well-known that collecting good statistics for these functions requires very large systems~\cite{karmakar2010analysis,berthier2007spontaneousa,berthier2007spontaneousb,flenner2013dynamic}. This represents a numerical effort on its own, which we leave for future work. 

We characterise growing dynamic lengthscales using an alternative method based on the chord length distribution. This method was introduced to analyse porous and bicontinuous structures~\cite{levitz1998off,testard2014intermittent,testard2011influence}. We note that the method would not faithfully characterise dynamic heterogeneities at temperatures above $T_{\rm mct}$, where relaxed domains have fuzzy shapes and ill-defined boundaries. It was however shown to efficiently determine characteristic lengthscales in compact bicontinuous structures~\cite{testard2014intermittent}, which are observed at $T_g< T < T_{\rm mct}$. Finally, this method does not \rev{suffer from considerations related to the choice of  statistical ensembles~\cite{berthier2007spontaneousa,berthier2007spontaneousb}.} 

\begin{figure}
    \includegraphics[width=\columnwidth]{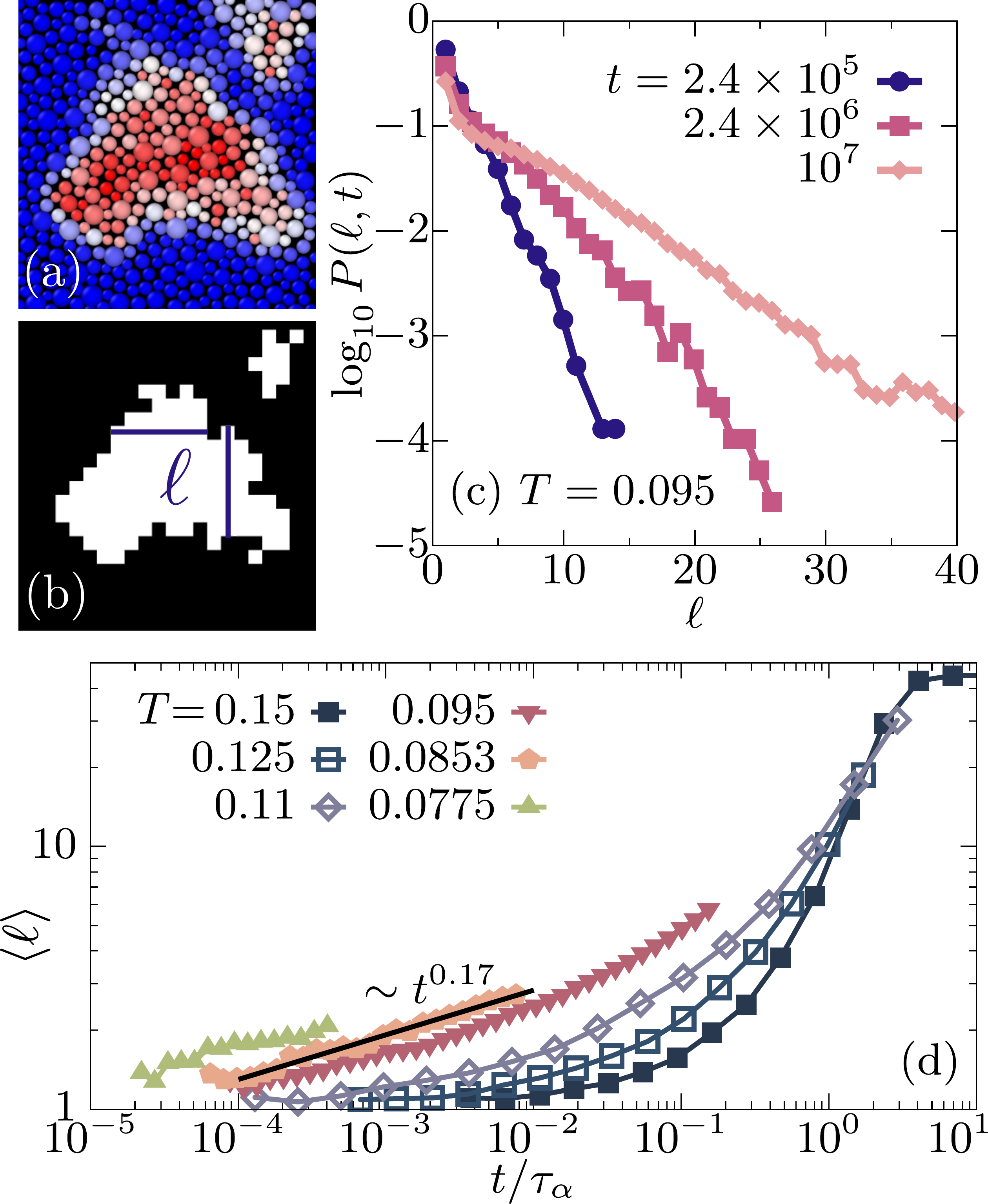}
    \caption{{\bf Slow coarsening of relaxed domains measured by chord length distributions in $2d$.} 
    (a) Snapshots of a portion of the two-dimensional liquid with a mobility field at $T=0.095$ (colour code from Fig.~\ref{fig:redblue}). (b) Discretised version of the mobility field, used to measure the length $\ell$ of chords drawn vertically and horizontally which intersect relaxed domains (white). (c) Chord length distribution $P(\ell,t)$ at $T=0.095$ and increasing times. (d) Growth of the average chord length $\langle \ell \rangle$ with time. A slow algebraic \rev{growth} develops at early times and low temperatures.}
\label{fig:plt}
\end{figure}

First, we identify relaxed regions via the bond-breaking correlation, see Fig.~\ref{fig:plt}(a). To determine the spatial extension of relaxed domains, it is convenient to perform a coarse-graining and binarise the $C_B^i$ field. To do so, we discretise the configuration into a grid of cells of linear size 1 and compute their average $C_B^i$, particles being weighted by their area overlap with each cell, which is then thresholded at 0.5 between mobile and quiescent cells. We illustrate in Fig.~\ref{fig:plt} how our procedure transforms a particle configuration (a) into a discrete lattice of mobile and quiescent cells (b). The binary grid is then used to measure the distribution of chord length $\ell$, defined as horizontal and vertical segments intersecting mobile domains, as shown in Fig.~\ref{fig:plt}(b). 

After averaging, the method yields the probability distribution of chord lengths $P(\ell,t)$ at any temperature $T \leq T_{\rm mct} $ and time $t$. As an example, we show in Fig.~\ref{fig:plt}(c) the chord length distribution measured at $T = 0.095$ in $2d$ for three different times. All distributions have a maximum at very small $\ell \approx 1$, presumably due to the roughness of the domains. At larger $\ell$, the distribution is smooth and decays over a typical distance which grows with increasing time. The decay at large $\ell$ is roughly exponential, as seen in other systems~\cite{testard2014intermittent}. 

To extract a time dependent lengthscale characterising the growing size of the domains of relaxed particles, we compute the first moment of the chord length distribution,
\begin{equation}
\langle \ell \rangle = \int_{0}^{\infty}\mathrm d  \ell P( \ell,t)  \ell .
\label{eq:Lt}
\end{equation}
The time dependence of $\langle \ell \rangle$ is shown in Fig.~\ref{fig:plt}(d) for a broad range of temperatures.

The time dependence of the typical chord length qualitatively changes with temperature. Close to $T_{\rm mct} = 0.12$ and above, the average relaxation time $\ta$ is not yet very large. Many small domains appear spontaneously across the system at early times, and they rapidly merge with one another. The merging of independent clusters is the mechanism that controls the growth of the typical domain size in the time regime near $\ta$. \rev{In other words, dynamic facilitation is present above $T_{\rm mct}$ but it does not seem to directly control the time dependence of spatially correlated dynamic domains.}  

At lower temperature instead, we observe a sparse population of regions that appear at early times and coarsen independently over a large time window before different domains start to merge. In this early time regime, which only exists at temperatures much lower than $T_{\rm mct}$, the typical domain size appears to grow as a power law, 
\begin{equation}
\langle \ell \rangle(t) \sim t^{1/z(T)},
\end{equation}
with a dynamic exponent $z(T)$ which increases slowly as $T$ decreases. For $T=0.0853$ we find for instance $z(T) \approx 5.9$ \rev{({\it i.e.}, $1/z \approx 0.17$).} This very large value of the dynamic exponent $z(T)$ implies that the growth of relaxed regions is strongly sub-diffusive in the time regime $t \ll \ta$. The decrease of this exponent with temperature demonstrates that, while the dynamics repeats itself with increasing probability, it also takes longer for mobility to spread to neighbouring regions. As a word of caution, we note that our numerical determination of $z(T)$ is performed over ranges of timescales and temperatures which are highly constrained, leaving open the possibility that the domain growth is actually better described by a logarithmic increase, for instance. Finally, the very slow growth of $\langle \ell \rangle$ in the $2d$ model echoes the similarly slow growth of the average cluster size $n(t)$ measured in the $3d$ model in Fig.~\ref{fig:clust_3d}. 

\subsection{\rev{How the story ends:} Lifetime of dynamic heterogeneity}
 
\label{sec:lifetime} 
 
We finally describe how structural relaxation takes place at very long times, beyond the structural relaxation time. Typically, $\ta$ is defined from the decay of a time correlation function $C(t)$ below an arbitrary level, for instance $C(\ta)=e^{-1}$. When $C(t)$ is the self-intermediate function this means that particles have, on average, moved by a distance $2 \pi / q$ after $\ta$. Of course, a similar value is reached if a fraction of the particles has moved a lot, while the rest is still immobile. Using the bond-breaking correlation function at half value, as we frequently did above, means that either all particles have on average lost half of their neighbours, or that one half of the particles has lost all of their neighbours \rev{while the others did nothing.} In other words, the definition of an `average' relaxation time $\ta$ does not imply that all particles have completely relaxed after a time $\ta$. In fact, the snapshots shown above clearly demonstrate that, at very low temperatures, a significant fraction of the particles have not relaxed at all after $\ta$ while the others have fully relaxed. This paragraph \rev{merely} reformulates more explicitly the well-established statement that the dynamics of supercooled liquids is spatially heterogeneous~\cite{ediger2000spatially}.  

\begin{figure}
   \includegraphics[width=0.9\columnwidth]{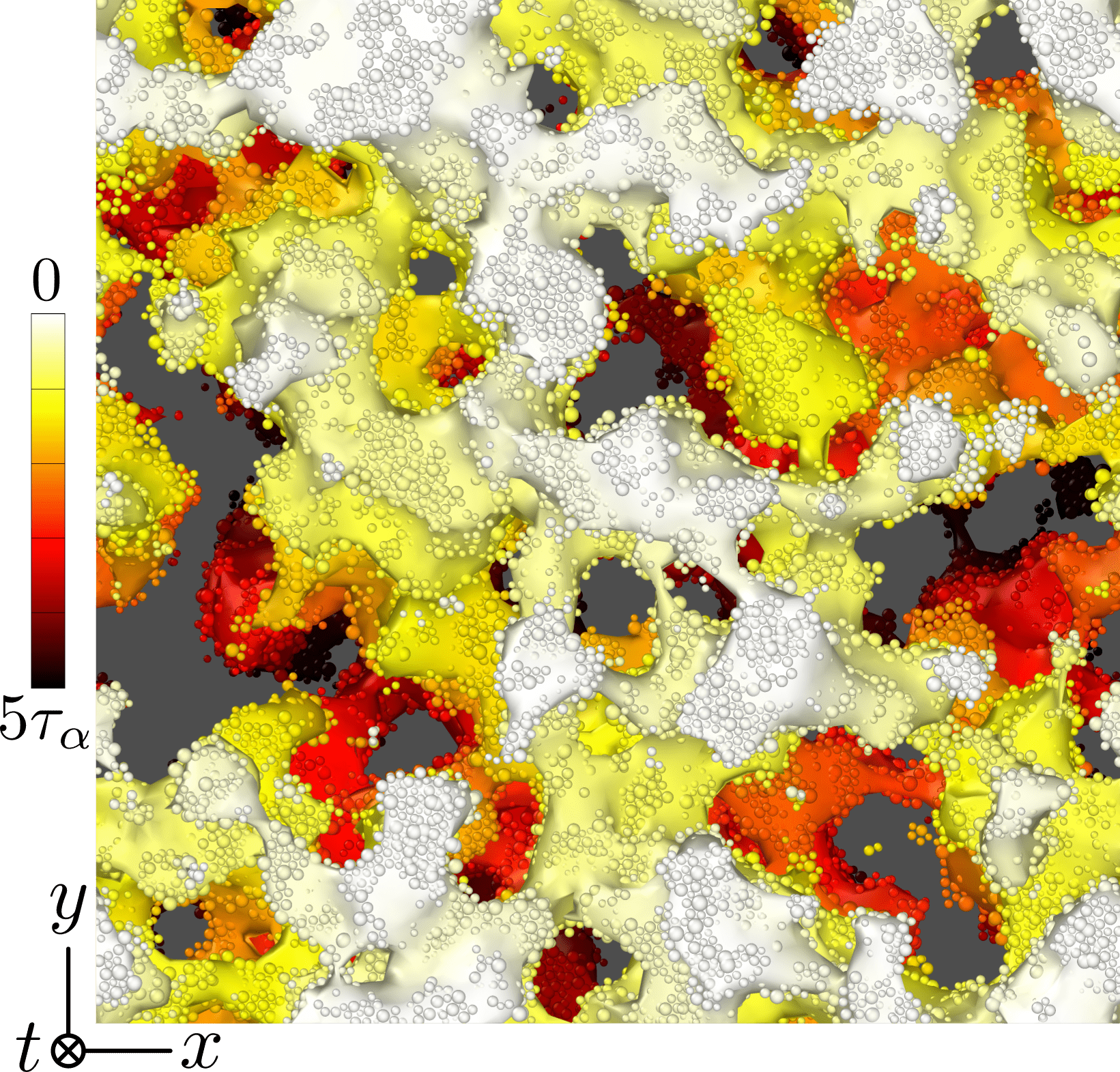}
   \caption{{\bf Relaxation of very slow domains in the $2d$ model.} $xy$ projection of the space-time mobility trajectory shown in Fig.~\ref{fig:bubbles}. The colour code indicates, in a linear scale, the time at which particles become mobile for the first time.}
   \label{fig:maplinear}
\end{figure}

We pushed the simulations at $T=0.09$ up to five times longer than the $\alpha$-relaxation time, and \rev{represent} the outcome in Fig.~\ref{fig:maplinear}. We show the $xy$ projection of the space-time representation of Fig.~\ref{fig:bubbles}, the time axis now pointing inside the plane, with a linear colour code from short (white) to long (dark) times. When accumulated over such a large timescale, we observe that a significant fraction of the system remains completely unrelaxed (grey domains). We also observe that the latest stages of the dynamics where some slow regions relax for the first time (red regions) spread from regions that have already relaxed. We have made similar observations in multiple trajectories: the slowest regions to relax are always invaded and relaxed from their boundaries. 

In other words, the regions that have already relaxed at $\ta$ are progressively invading the ones that relax slower than average. This suggests that the tail of the correlation function at very long times is physically controlled by the slow coarsening process described in the previous section. 

This has two important consequences. 

First, this implies that the grey unrelaxed regions in Fig.~\ref{fig:maplinear} are extremely stable and would tend to relax over a timescale that is much larger than the average relaxation time $\ta$. However, their environment is not as stable and the relaxation which takes place there facilitates or accelerates the relaxation of these slow domains. In Ref.~\cite{scalliet2021excess} we argued that this phenomenon provides a physical explanation for the asymmetric shape of relaxation spectra in deeply supercooled liquids in terms of an underlying distribution of `natural' relaxation times that has a very broad tail that is cutoff by the facilitated relaxation of the slowest regions. This mechanism has been discussed in various contexts~\cite{palmer1984models,xia2001microscopic,rehwald2012how,bhattacharyya2008facilitation,berthier2019can}. 

Second, the timescale over which the slowest regions relax in the system controls the so-called lifetime $\tau_{\rm dh}$ of the dynamic heterogeneity. Physically, a bicontinuous pattern of the mobility field with fast and slow particles emerges near $\ta$, as shown for instance in Fig.~\ref{fig:tempevolution2d}. If the relaxation time in the slow regions in such plots were given by a timescale $\tau_{\rm dh} \gg \ta$, then the same mobility pattern would be found over many consecutive time intervals of duration $\ta$, and would only start to change significantly after a time $\tau_{\rm dh}$~\cite{heuer1997information}. The study of this lifetime has been the subject of intense experimental~\cite{ediger2000spatially,schmidt1991nature,wang1999long,paeng2015ideal,paeng2016single} and numerical~\cite{leonard2005lifetime,kim2010multitime,kim2013multiple,berthier2021self} studies. 

Our simulations show that, as a result of dynamic facilitation, the lifetime of the slow regions is controlled by a combination of two factors: (i) the relaxation time of the fast regions; (ii) the coarsening of mobility from one relaxed region to the next. Because the latter depends very weakly on temperature, we expect that $\tau_{\rm dh}$ may increase slightly more slowly than $\ta$ but should not be strongly decoupled from it. In simulations where complete decorrelation can be observed, we find that after 20-40 times $\ta$, the entire system has completely relaxed and information about dynamic heterogeneity is then totally lost. This conclusion appears consistent with recent experiments~\cite{paeng2015ideal,paeng2016single} and simulations~\cite{berthier2021self}. \rev{Such a long timescales also justifies our statement about equilibration in supercooled liquids, which cannot be reached before at least 20-40$\,\tau_\alpha$ as otherwise some particles would not have relaxed at all.} 

\section{Discussion}

\label{sec:discussion}

\subsection{Summary of main novel results}

Let us first briefly recapitulate the main new results arising from studying dynamics over long timescales at very low temperatures.

First, we observed that in the very low temperature regime the dynamics starts at sparse locations and takes the form of localised but quite complex relaxation events. Detailed analysis reveals the emergence of power laws characterising the time dependence of three quantities: the number of independent clusters of mobile particles, the distribution of waiting time for their appearance, and the high-frequency dependence of ensemble-averaged relaxation spectra (excess wing). We demonstrated that these power laws are intimately connected, and their amplitude strongly depends on the stability of the system.  

Second, we described how these early relaxation events induce an accumulation of relaxation events repeatedly taking place at roughly the same location leading to the slow coarsening of the relaxed domains. These observations account for the emergence, at a coarse-grained level, of dynamic facilitation. In particular, we characterised a dynamic exponent $z(T)$ relating timescales and lengthscales, which we extracted by introducing a chord length analysis of the dynamic heterogeneity. The corresponding growth of relaxed regions is very slow and strongly sub-diffusive with a large value of $z(T)$, possibly logarithmic.  

Third, we observed a qualitative evolution of the nature of the spatially heterogeneous dynamics at all timescales as temperature is decreased much below the mode-coupling crossover. In particular, we found an increasing segregation between mobile and immobile regions resulting in spatial fluctuations of the mobility that become more compact with smoother boundaries at low temperatures.

Fourth, we found that \rev{at low temperature} the regions that relax the slowest at timescales much larger than $\ta$ seem so stable that the fastest mechanism to relax them is via the propagation of mobility from faster regions surrounding them, that is, via dynamic facilitation. \rev{Our simulations thus demonstrate that the lifetime of the dynamic heterogeneity is controlled by dynamic facilitation.}

\subsection{Comparison with previous simulations}

Computational research regarding the dynamics of supercooled liquids is immense and covers several decades of work~\cite{barrat2022computer}. Because we explore a temperature regime that was not accessible to simulations before, it is useful to contrast our main results with earlier studies. This is organised in three broad topics.

First, we discuss the emergence of excess wings associated to short-time localised relaxation events characterised by a broad waiting time distribution, as reported in Sec.~\ref{sec:clusters}. This finding more broadly refers to the topic of secondary relaxations, or $\beta$-processes, in the dynamics of supercooled liquids~\cite{doi:10.1063/1.1286035,stevenson2010universal}. We recall that excess wings appear in experiments in a time window roughly between $1$~$\mu$s and 1~s, for temperatures much lower than $T_{\rm mct}$~\cite{nagel_scaling,nagel_scaling2,kudlik1998slow}. Simulations performed without the swap Monte Carlo algorithm cannot reach equilibrium at temperatures where excess wings appear. As a consequence, earlier attempts to explore similar time and temperature windows necessarily dealt with non-equilibrium glasses obtained by crossing a computer glass transition at a temperature much higher than $T_g$, typically near $T_{\rm mct}$. In Sec.~\ref{sec:suppression} we showed that relaxation events detected in such poorly annealed glasses are not representative of those observed in equilibrium materials at the same temperature. In a recent series of simulations using computer models not very different from ours~\cite{yu2018fundamental,yu2017structural}, these events were mathematically described as an additive secondary peak in relaxation spectra. Other works have reported similar conclusions regarding the existence of a $\beta$-process~\cite{zhang2021fast}. Our results show that this description may not be adequate as the high-frequency shoulder observed in these works in fact turns into extended power laws in equilibrium at lower temperatures.

Our work also sheds light on an important issue regarding secondary relaxations. A number of numerical studies have reported the existence of a $\beta$-process taking the form of a peak in the frequency domain located at a frequency fully decoupled from $\ta$. These studies analysed the relaxation behaviour of particles that are more complex than the point particles studied here, for instance polymeric systems~\cite{bedrov2005molecular,bedrov2011secondary} or particles with shapes~\cite{roland20212molecular,roland2018cooperativity,shiraishi2022johari}. These studies suggest that intramolecular degrees of freedom can make relaxation spectra even more complex than the ones we report. However, our observation that excess wings taking the form of extended power laws exist in our much simpler model shows that secondary processes do exist in the absence of intramolecular degrees of freedom. 

A second major theme in computer studies is the idea that a crossover temperature separates two physical regimes in the dynamics of supercooled liquids~\cite{kob1999computer}. This is theoretically rationalised by the existence of a critical temperature controlling the dynamics of supercooled liquids in the context of mean-field approaches~\cite{kirkpatrick1987p,gotze2009complex,parisi2020theory}. This critical temperature is similar to the one predicted from mode-coupling theory although its interpretation as an artefact of mean-field approximations has only recently been fully clarified~\cite{ikeda2012mode,maimbourg2016solution}. It is generally agreed that such mean-field approaches may usefully describe the first few decades of the dynamic slowdown, which is precisely the temperature regime studied over the years by computer simulations~\cite{kob1999computer}. Having access to considerably lower temperatures confirms that relaxation dynamics keeps changing with decreasing temperatures. \rev{However, the present results are in fact more naturally described as a progressive evolution towards low-temperature physics, with no sharp change at a specific temperature.} Importantly, most of the features that we report only become prominent and unambiguous at the lowest temperatures that we can analyse. We suggest that a considerable amount of work is needed to reassess conclusions drawn in previous works \rev{at higher temperatures} .     

Third, our results show how dynamic facilitation emerges at very low temperature and plays an important role in both the approach to $\ta$ and at longer times. The concept of dynamic facilitation has a long history in the glass literature~\cite{palmer1984models,fredrickson1984kinetic}, to the point that it serves as a basis for the construction of a large family of kinetic glass models~\cite{ritort2003glassy}, which can display many phenomena observed in more realistic particle systems and have been in particular very useful to characterise dynamic heterogeneity~\cite{garrahan2011kinetically}. 

Direct signatures of dynamic facilitation in molecular dynamics simulations have been discussed before~\cite{gebremichael2004particle,vogel2004spatially,bergroth2005examination,candelier2010spatiotemporal,keys2011excitations}. The most direct studies have in particular followed similar principles: first use some thresholding procedure to detect the time at which a given particle is moving ({\it i.e.}, performing some jump), and then search for the enhanced probability that a nearby particle will relax not too far in the future~\cite{vogel2004spatially}. Our philosophy has been conceptually different. We did not perform single particle analysis of jump dynamics but rather quantified how the observed collective relaxation events effectively give rise to dynamic facilitation when observed over some large timescales and lengthscales, as in the space time representation of Fig.~\ref{fig:bubbles} or via the emergence of a dynamic exponent $z(T)$. Our simulations suggest that dynamic facilitation becomes more prominent and should be more easily characterised at lower temperatures, which seems to contradict earlier findings~\cite{candelier2009building}. 

Finally, a recent numerical study suggests that dynamic facilitation may become long-ranged below $T_{\rm mct}$ and may result from elastic interactions~\cite{chacko2021elastoplasticity}. We have not repeated the detailed analysis proposed in this work at the much lower temperatures analysed here, but the gradual coarsening of relaxed domains suggests that facilitation does not act over very large distances and can be quantitatively described, at a coarse-grained level, by a dynamic exponent similar to the one emerging from nearest neighbour facilitation in kinetically constrained models. Reconciling these two observations represents a worthwhile topic for future work.

\subsection{Consequences for glass transition physics}

What do we learn about the physics of supercooled liquids, and how does this knowledge impact existing theories of the glass transition?

An important lesson drawn from our results is that equilibrium relaxation dynamics at low temperatures appears qualitatively different from observations made in the usual temperature regime covered by molecular dynamics studies. Whereas particle motion results from collective mode-like mobility patterns involving most particles at high $T$, we observe instead localised particle motion that completely relax the structure of mobile particles leaving the rest of the system essentially unaffected at low $T$. These observations capture, in real space, the commonly accepted idea that dynamics smoothly transitions from a non-activated relaxation regime that efficiently relaxes the system at high temperature to a regime dominated by activated processes involving a finite number of particles at low temperature \rev{embedded in an otherwise frozen elastic matrix.}   

\begin{figure}
   \includegraphics[width=\columnwidth]{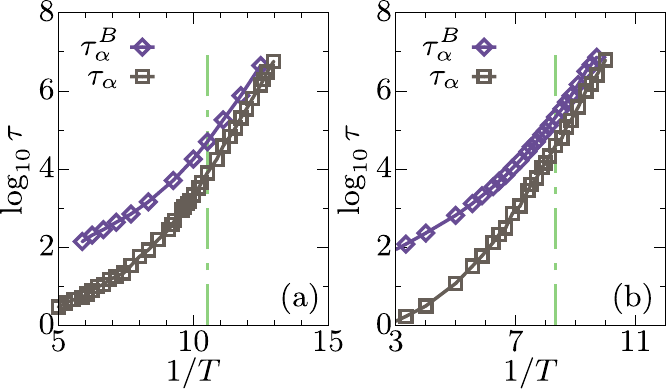} 
   \caption{{\bf Convergence of relaxation times at low temperature in $3d$ (a) and $2d$ (b)}. Relaxation times $\tau_\alpha$ defined from the self-intermediate scattering function $F_s(t)$ (a) and bond-orientational correlation $C_\Psi(t)$ (b), compared to the relaxation time $\tau_\alpha^B$ defined from the bond-breaking correlation $C_B(t)$. The very different timescales measured at high temperatures become identical at low temperatures, much below $T_{\rm mct}$ (dotted dashed line).}
          \label{fig:convergence}
\end{figure}

In practice, this crossover happens quite gradually as the relaxation time increases by about seven orders of magnitude, and it is difficult to claim that a particular temperature scale, such as the mode-coupling crossover $T_{\rm mct}$, marks a sharp change in the physics. Previous studies offered examples of both sharp~\cite{broderix2000energy,grigera2002geometric,coslovich2019localization} and smooth~\cite{doliwa2003what,berthier2012finite,kob2012non,coslovich2018dynamic,das2022crossover} changes between the two types of physics, and our analysis of the relaxation dynamics is in better agreement with the latter family of studies. 

To corroborate this point, we present in Fig.~\ref{fig:convergence} the temperature evolution of the average relaxation times deduced from two types of correlation functions. Previous simulations performed in the non-activated high-temperature regime have reported a strong decoupling~\cite{yamamoto1998dynamics, shiba2012relationship, flenner2019viscoelastic, kawasaki2017identifying} between the bond-breaking correlation $C_B(t)$ and more conventional time correlation functions, such as the self-intermediate scattering function $F_s(t)$ or the bond-orientational correlation $C_\Psi(t)$. This has been rationalised by the presence of vibrational mode-like mobility patterns such as the one shown in Fig.~\ref{fig:smallclusters}, that are able to decorrelate $F_s(t)$ and $C_\Psi(t)$, but would leave $C_B(t)$ essentially unaffected~\cite{shiba2012relationship}. The results shown in Fig.~\ref{fig:convergence} confirm this finding at high temperatures. However, pushing the analysis to much lower temperatures reveals that these two timescales eventually converge. At the lowest temperature, localised and complex activated relaxation processes result in large displacements that are able to simultaneously decorrelate $F_s(t)$ or $C_\Psi(t)$ and $C_B(t)$. This qualitative change in the dynamics towards localised activated relaxation events emerges gradually, and the two relaxation times start to coincide at temperatures that are in fact much lower than $T_{\rm mct}$. \rev{In practice this implies that the dynamic heterogeneity at these low temperatures becomes fairly independent of the particular observable and specific choices made to detect particle mobility.}

The idea of a crossover towards activated dynamics at low temperatures has a long history, and is in particular at the core of the random first order transition (RFOT) theory~\cite{kirkpatrick1989scaling,lubchenko2003barrier,gotze2009complex}. The theoretical explanation is that the dynamic transition predicted in the mean-field limit must be avoided in finite dimensions, as the lifetime of glassy states necessarily becomes finite~\cite{parisi2020theory}. A crossover between non-activated high-temperature dynamics and activated localised events also emerges naturally in kinetically constrained models~\cite{berthier2003real}, but this occurs without invoking an avoided dynamic transition. Therefore the observation of a crossover, in itself, does not confirm any particular theoretical approach. 

More interestingly perhaps, the direct observation of localised activated events reported here suggests that one should now be able to understand and characterise better their nature. Our preliminary investigations show that relaxation dynamics at low temperatures involves an extremely large number of transitions between inherent structures for instance, implying that the potential energy landscape cannot be used to gather information about relaxation dynamics as already noticed~\cite{heuer2008exploring}. It remains numerically challenging to unambiguously group inherent structures into larger metabasins~\cite{buchner2000metastable, doliwa2003what, doliwa2003hopping, doliwa2003energy, vogel2004particle} or to quantify free energy barriers~\cite{baity2021revisiting}, as envisioned within RFOT theory. Revisiting these earlier attempts is an important goal to characterise the nature of activated dynamics in the regime where it is effectively present.

We have repeatedly argued that dynamic facilitation emerges in the relaxation dynamics at low temperatures. This observation is a crucial new piece of evidence in the \rev{long-lasting debate} between dynamic and static explanations of the glass transition~\cite{tarjus2011overview}. Clearly, the thermodynamic RFOT theory picture of a liquid broken into a mosaic of droplets undergoing collective activated relaxation events~\cite{kirkpatrick1989scaling,bouchaud2004on} is invalidated by our results, \rev{at least in the simplest picture of independently relaxing droplets.}

There have been several attempts to introduce some degree of dynamic facilitation within RFOT theory, with the argument that fast relaxing droplets could affect the dynamics of slower ones in their neighbourhood, as described phenomenologically in Refs.~\cite{xia2001microscopic,berthier2019can}, or more formally in Refs.~\cite{bhattacharyya2005bridging,bhattacharyya2008facilitation}.
\rev{Our results show that such combination would be needed to account for the dynamics in laboratory experiments, and quantitatively explain the temperature evolution of central physical quantities, such as the $\alpha$-relaxation time or the length scale of dynamic heterogeneity~\cite{biroli2022rfot}.}

\begin{figure}
   \includegraphics[width=\columnwidth]{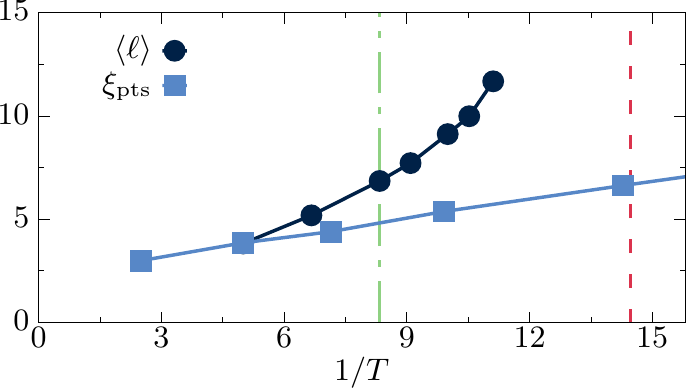}
   \caption{{\bf Decoupling of static and dynamic lengthscales in $2d$ supercooled liquids}. The dynamic lengthscale is the average chord length $\langle \ell \rangle$ of dynamic heterogeneities at $C_B=0.5$. The point-to-set lengthscale $\xi_{\rm pts}$ is equal to $2R_c$, with $R_c$ the crossover radius from Ref.~\cite{berthier2019zero}. Mode-coupling crossover temperature (dashed-dotted) and glass transition temperature (dashed) are indicated.}
          \label{fig:divergence}
\end{figure}

In Fig.~\ref{fig:divergence} we provide a decisive quantitative support for this conclusion in $2d$. We directly compare the temperature evolution of the static point-to-set length $\xi_{\rm pts}$~\cite{montanari2006rigorous} that sets the typical size of the droplets in the mosaic picture~\cite{kirkpatrick1989scaling,bouchaud2004on}, and the typical lengthscale of the dynamic heterogeneity at the structural relaxation time deduced from the chord length analysis described in Sec.~\ref{sec:coarsening}. The data for $\xi_{\rm pts}$ are taken from Ref.~\cite{berthier2019zero} for our $2d$ model. A strong decoupling between static and dynamic lengthscales is observed. This has been reported before for $T > T_{\rm mct}$~\cite{kob2012non,charbonneau2013decorrelation}, in a regime where activated dynamics should not be relevant. This had left open the possibility that the static point-to-set length could catch up with the dynamic one in a low temperature regime dominated by activated processes, as discussed in Ref.~\cite{kob2012spatial}. Our results invalidate this hypothesis, and demonstrate that, at least for our $2d$ system, the characteristic size of the dynamic heterogeneity is not controlled by $\xi_{\rm pts}$, even at temperatures near the experimental glass transition $T_g$. This is a major conclusion that emerges from the present study: the simplest formulation of RFOT theory does not describe well our data without invoking an important facilitation component~\cite{xia2001microscopic,bhattacharyya2008facilitation,berthier2019can}. 

Does this conclusion invalidate thermodynamic theories of the glass transition? Clearly not. First, future work should consolidate our findings across a broader range of models, and, more importantly, in $3d$ where the measurement and comparison of various lengthscales require a massive computational effort. Establishing whether $3d$ models behave as in Fig.~\ref{fig:divergence} is an important task. Second, it could still be that the emergence of clusters at short times that triggers further relaxation over growing domains is due to the type of activated dynamics envisioned by RFOT theory and these events could very well be controlled by the growing static lengthscale $\xi_{\rm pts}$. As an illustration of this idea, we have marked in Fig.~\ref{fig:tempevolution2d} the spatial extent of the point-to-set lengthscale, which compares reasonably well with the spatial extension of the domains which first relax. If correct, this interpretation would imply that structural relaxation emerges from a combination of physical ingredients, requiring descriptive tools stemming from both dynamic and static approaches~\cite{bhattacharyya2008facilitation}.
\rev{More broadly, future work should reconcile the present findings regarding the dynamics of supercooled liquids to the recent detailed characterisation of global~\cite{PhysRevE.102.042129,guiselin2022statistical} and local~\cite{berthier2021self,doi:10.1063/5.0086517} thermodynamic fluctuations of the overlap order parameter~\cite{guiselin2020overlap, isglass} which are at the root of RFOT theory.} 
Third, the accumulation of relaxation events over long periods of time in certain locations suggests that dynamic facilitation necessarily has a static origin~\cite{ortlieb2021relaxation,ganapathi2022structural}. In addition, explaining how relaxation at one place triggers relaxation in the neighbourhood should also require linking statics to dynamics~\cite{keys2011excitations}. Therefore, the observation of dynamic facilitation does not automatically imply that static considerations and structural and thermodynamic information become irrelevant. Future work should explore how to best describe or incorporate dynamic facilitation effects in the context of RFOT theory.  

Does the observation of dynamic facilitation automatically validate all predictions stemming from analogies with kinetically constrained models~\cite{chandler2010dynamics}? Clearly not, and a lot remains to be done. A key result we obtained in this regard is the direct measurement of a large dynamic exponent $z(T)$. The measured value invalidates entire families of kinetically constrained models, including the simplest version of the Fredrickson-Andersen facilitated model~\cite{whitelam2005renormalization}, but also the large family of collective kinetically constrained models~\cite{fredrickson1984kinetic,fredrickson1985facilitated,kob1993kinetic,toninelli2006jamming,toninelli2008new}, where non-trivial defects display diffusive motion~\cite{toninelli2004spatial}, corresponding to $z=2$. Clearly our results are much closer to the behaviour found in the family of anisotropic East~\cite{jackle1991hierarchically,berthier2005numerical} and arrow~\cite{garrahan2003coarse} models where defects are subdiffusive~\cite{garrahan2011kinetically,sollich1999glassy,sollich2003glassy}. However, further work should establish for instance whether the power laws characterising the short time dynamics revealed in Sec.~\ref{sec:clusters} are similar to the ones reported in Ref.~\cite{berthier2005numerical} in the $3d$ East model which also give rise to excess wings. It would also be useful to understand whether and how the concept of kinetic constraints emerges from \rev{microscopic atomic motion and particle interaction,} and whether the exponent $z(T)$ varies quantitatively as predicted in the East model. 

\subsection{Further perspectives}

Finally we describe four important research directions suggested by our results. 

The first important task ahead of us is to test whether the results obtained here can be generalised to other models, by changing the type of particle interactions, and possibly generalising to more complicated particle shapes and molecules. This clearly requires improving the swap Monte Carlo algorithm even further~\cite{berthier2019efficient,parmar2020stable} as well as developing molecular dynamics softwares and \rev{hardware} that are even more efficient~\cite{plimpton1995fast,bailey2017rumd}. It would also be useful to vary the spatial dimensions over a broader spectrum~\cite{eaves2009spatial,adhikari2021spatial,berthier2020finite}, to see whether glassy dynamics changes qualitatively by increasing $d$ over a larger range.

A second important task concerns a more detailed characterisation of the spatially heterogeneous dynamics at the very low temperatures that can now be accessed. Decades of work have led to the development of an arsenal of tools consisting of multi-point correlation functions and dynamic susceptibilities~\cite{toninelli2005dynamical,berthier2011dynamical}. These quantities are well-known and their physical content fully understood. Technically however, there are subtleties and potential artefacts that need to be carefully considered~\cite{berthier2007spontaneousa}. In addition, a proper measurement of correlation lengthscales, their time and temperature evolution, as well as the characterisation of the geometry of these correlations require significant statistics along with simulation boxes that are large enough~\cite{karmakar2010analysis,flenner2013dynamic}. Such measurements would allow a quantitative characterisation of the growth of characteristic lengthscales for dynamic heterogeneity as well as the evolution of the geometry of correlated regions for which quantitative scenarios exist~\cite{stevenson2006shapes,berthier2005numerical}. This is a clear task for future work, but it necessitates a large investment of computational resources. 

A third research effort should aim at a better characterisation of dynamic facilitation, which becomes more prominent at low temperatures than it is above $T_{\rm mct}$. It could be useful to explore the behaviour of the various tools introduced before~\cite{vogel2004particle,bergroth2005examination,keys2011excitations} to correlate single particle motion in space and time at much lower temperatures. Starting with Ref.~\cite{keys2011excitations}, several papers have developed computational tools to identify the analog of the excitations that define kinetically constrained models in particle-based simulations~\cite{speck2012constrained,isobe2016applicability,campo2020dynamical,ortlieb2021relaxation}. Would these algorithms be able to detect the sparse population of relaxing clusters that drives the relaxation near $T_g$? Methods should also be developed to more directly and more precisely measure the dynamic exponent $z(T)$ relating timescales and lengthscales, as this is an important emerging consequence of dynamic facilitation. Finally, we have merely observed the emergence of dynamic facilitation, but we did not provide a microscopic understanding of its origin in terms of structure, thermodynamics or geometry of the supercooled liquid. This ambitious task would be required to provide a fully microscopic picture of structural relaxation near $T_g$. 

A fourth ambitious goal is the development of more efficient molecular dynamics schemes that would allow us to fill the remaining gap between the 30 ms studied here, and the 100 s timescale that characterise most experiments. Our results show that the physics keeps changing qualitatively when shifting the `glass ceiling' by several orders of magnitude. For the type of models studied here, the remaining gap in timescales is about $10^4$. Such a colossal gap cannot be filled by hardware improvements or by performing longer simulations and will require creative development and invention of molecular dynamics techniques, maybe coupled to smart Monte Carlo moves. Filling this gap appears as the last obstacle before reaching the Holy Grail: simulating 100~s of the life of a supercooled liquid at the experimental glass transition. 

\section*{Author contribution}

C. Scalliet and B. Guiselin contributed equally to this work.

\acknowledgments
We would like to thank Jean-Philippe Bouchaud for his support, questions, and constant interest over the last two years. Many members of the Simons Collaboration `Cracking the glass problem' have challenged our numerical results and interpretations over the last two years, which helped us improve our manuscript. \rev{We thank G. Jung, D. S. Simmons, and P. G. Wolynes
for comments on our manuscript.} This work was supported by a grant from the Simons Foundation (\#454933, LB) and the European Research Council under the EU's Horizon 2020 Program, Grant No. 740269. CS acknowledges support from the Fondation l'Or\'eal for a L'Or\'eal-UNESCO For Women in Science Fellowship, a Herchel Smith Fellowship, University of Cambridge, and a Ramon Jenkins Research Fellowship from Sidney Sussex College, Cambridge. BG acknowledges support by Capital Fund Management - Fondation pour la Recherche.

\section*{Dedication}

\hfill {\it Sail to the moon.}
We would like to dedicate this article to Jean-Philippe Bouchaud on the occasion of his 60th birthday. Among many other reasons, Jean-Philippe's support, questions, and constant interest over the last two years helped us keep the fire burning.

\bibliography{longwings.bib}

\end{document}